\documentclass[prx, 10pt, twocolumn, floatfix, superscriptaddress] {revtex4-1}

\usepackage{times, mathrsfs, amsmath, amsfonts, graphics, graphicx, cancel, color, theorem, bbm, mathtools, amssymb, multirow}

\usepackage[dvipsnames]{xcolor}
\usepackage[english]{babel}
\usepackage[margin=.75in]{geometry}
\usepackage[T1]{fontenc}

\usepackage{xfrac,relsize,float,tikz}

\usepackage[unicode=true, breaklinks=false, pdfborder={0 0 1}, backref=false, colorlinks=true, linkcolor=blue, citecolor=blue]{hyperref}

\newcommand\copyrighttext{%
  \footnotesize Accepted in Physical Review Research 3 April, 2020. Click title to verify. Published under CC-BY 4.0.
 }
\newcommand\copyrightnotice{%
\begin{tikzpicture}[remember picture,overlay]
\node[anchor=south,yshift=10pt] at (current page.south) {\fbox{\parbox{\dimexpr\textwidth-\fboxsep-\fboxrule\relax}{\copyrighttext}}};
\end{tikzpicture}%
}

\newcommand{\eM}{\ensuremath{\epsilon\text{-machine}}}
\newcommand{\eMs}{\ensuremath{\epsilon\text{-machines}}}
\newcommand{\DCmu}{\ensuremath{\Delta C_{\mu}}}
\newcommand{\cev}[1]{\reflectbox{\ensuremath{\vec{\reflectbox{\ensuremath{#1}}}}}} 
\renewcommand{\v}[1]{\ensuremath{\mathbf{#1}}} 
\newcommand{\gv}[1]{\ensuremath{\mbox{\boldmath$ #1 $}}} 

\begin{document}
\copyrightnotice

\title{\href{https://doi.org/10.1103/PhysRevResearch.2.023219}{General anaesthesia reduces complexity and temporal asymmetry of the informational structures derived from neural recordings in \textit{Drosophila}}}

\author{Roberto N. Mu{\~n}oz}
\email{roberto.munoz@monash.edu}
\affiliation{School of Physics \& Astronomy, Monash University, Clayton, Victoria 3800, Australia}

\author{Angus Leung}
\email{angus.leung1@monash.edu}
\affiliation{School of Psychological Sciences, Monash University, Clayton, Victoria 3800, Australia}

\author{Aidan Zecevik}
\email{aidanzecevik@gmail.com}
\affiliation{School of Physics \& Astronomy, Monash University, Clayton, Victoria 3800, Australia}

\author{Felix A. Pollock}
\email{felix.pollock@monash.edu}
\affiliation{School of Physics \& Astronomy, Monash University, Clayton, Victoria 3800, Australia}

\author{Dror~Cohen}
\email{dror.cohen@nict.go.jp}
\affiliation{Center for Information and Neural Networks (CiNet), National Institute of Information and Communications Technology (NICT), Suita, Osaka 565-0871, Japan}
\affiliation{School of Psychological Sciences and Turner Institute for Brain and Mental Health, Monash University, Melbourne, Victoria 3800, Australia}

\author{Bruno van~Swinderen}
\email{b.vanswinderen@uq.edu.au}
\affiliation{Queensland Brain Institute, The University of Queensland, St Lucia, Queensland 4072, Australia}

\author{Naotsugu Tsuchiya}
\email{naotsugu.tsuchiya@monash.edu}
\affiliation{School of Psychological Sciences and Turner Institute for Brain and Mental Health, Monash University, Melbourne, Victoria 3800, Australia}
\affiliation{Advanced Telecommunications Research Computational Neuroscience Laboratories, 2-2-2 Hikaridai, Seika-cho, Soraku-gun, Kyoto 619-0288, Japan}
\affiliation{Center for Information and Neural Networks (CiNet), National Institute of Information and Communications Technology (NICT), Suita, Osaka 565-0871, Japan}

\author{Kavan Modi}
\email{kavan.modi@monash.edu}
\affiliation{School of Physics \& Astronomy, Monash University, Clayton, Victoria 3800, Australia}

\date{\today}
\begin{abstract}
We apply techniques from the field of computational mechanics to evaluate the statistical complexity of neural recording data from fruit flies. First, we connect statistical complexity to the flies' level of conscious arousal, which is manipulated by general anaesthesia (isoflurane). We show that the complexity of even single channel time series data decreases under anaesthesia. The observed difference in complexity between the two states of conscious arousal increases as higher orders of temporal correlations are taken into account. We then go on to show that, in addition to reducing complexity, anaesthesia also modulates the informational structure between the forward and reverse-time neural signals. Specifically, using three distinct notions of temporal asymmetry we show that anaesthesia reduces temporal asymmetry on information-theoretic and information-geometric grounds. In contrast to prior work, our results show that: (1) Complexity differences can emerge at very short time scales and across broad regions of the fly brain, thus heralding the macroscopic state of anaesthesia in a previously unforeseen manner, and (2) that general anaesthesia also modulates the temporal asymmetry of neural signals. Together, our results demonstrate that anaesthetised brains become both less structured and more reversible.
\end{abstract}

\maketitle

\section{Introduction}  
\label{sec:Intro}

Complex phenomena are everywhere in the physical world. Typically, these emerge from simple interactions among elements in a network, such as atoms making up molecules or organisms in a society. Despite their diversity, it is possible to approach these subjects with a common set of tools, using numerical and statistical techniques to relate microscopic details to emergent macroscopic properties~\cite{Thurner2018}. There has long been a trend of applying these tools to the brain, the archetypal complex system, and much of neuroscience is concerned with relating electrical activity in networks of neurons to psychological and cognitive phenomena~\cite{CognitiveNeurosciences}. In particular, there is a growing body of experimental evidence~\cite{Boly2013}, that neural firing patterns can be strongly related to the level of conscious arousal in animals. 

In humans, level of consciousness varies from very low in coma and under deep general anaesthesia, to very high in fully wakeful states of conscious arousal~\cite{Laureys2012}. With the current technology, precise discrimination between unconscious vegetative states and minimally conscious states are particularly challenging and remains a clinical challenge~\cite{NeurologyOfConsciousness}. Therefore, substantial improvement in accuracy of determining such conscious states using neural recording data will have significant societal impacts. Towards such a goal, neural data has been analysed using various techniques and notions of \textit{complexity} to try to find the most reliable measure of consciousness~\cite{Engemann2018, Sitt2014}.

One of the most successful techniques to date in distinguishing levels of conscious arousal is the \textit{perturbational complexity index}~\cite{Massimini2005, Casali2013, Casarotto2016}, which measures the neural activity patterns that follows a perturbation of the brain through magnetic stimulation. The evoked patterns are processed through a pipeline then finally summarised using Lempel-Ziv complexity~\cite{Casali2013}. This method is inspired by a theory of consciousness, called \textit{integrated information theory} (\textbf{IIT})~\cite{Tononi2004, Tononi2016}, which proposes that a high level of conscious arousal should be correlated with the amount of so-called \textit{integrated information}, or the degree of differentiated integration in a neural system (see Ref.~\cite{Oizumi2014} for details). While there are various ways to capture this essential concept~\cite{Mediano2019, Barrett2011}, one way to interpret integrated information is as the amount of loss of information a system has on its own future or past states based on its current state, when the system is minimally disconnected~\cite{Tegmark2016, Oizumi2016PNAS, Oizumi2016PLOS}.

These complexity measures, inspired by IIT, are motivated by the fundamental properties of conscious phenomenology, such as informativeness and integratedness of any experience \cite{Tononi2004}. While there are ongoing efforts to accurately translate these phenomenological properties into mathematical postulates~\cite{Oizumi2014}, such translation often contains assumptions about the underlying process which are not necessarily borne out in reality. For example, the derived mathematical postulates in IIT assume Markovian dynamics, i.e., that the future evolution of a neural system is determined statistically by its present state~\cite{Barrett2011}. Moreover, IIT requires computing the correlations across all possible partitions between subsystems, which is computationally heavy~\cite{Tegmark2016} in relation to methods which do not require such partitioning to work. Assuming that the hierarchical causal influences in the brain would manifest as oscillations across a range of frequencies and spatial regions~\cite{Buzsaki2006}, non-Markovian temporal correlations likely play a significant role in explaining any experimentally measurable behaviours, including the level of conscious arousal. There is therefore, scope for applying more general notions of complexity to meaningfully distinguish macroscopic brain states that support consciousness.

A conceptually simple approach to quantifying the complexity of time series data, such as the fluctuating potential in a neuron, is to construct the minimal model which statistically reproduces it. Remarkably, this minimal model, known as an \textit{epsilon machine} (\eM{}), can be found via a systematic procedure  which has been developed within the field of computational mechanics~\cite{CrutchPRL1989, epsilonMachines2, CrutcharXiv2017}. Crucially, \eMs{} account for multiple temporal correlations contained in the data and can be used to quantify the \emph{statistical complexity} of a process -- the minimal amount of information required to specify its state. As such they have been applied over various fields, ranging from neuroscience~\cite{Haslinger2009, Klinkner2006} and psychology~\cite{CSSR2} to crystallography~\cite{Varn2004} and ecology~\cite{Boschetti2008}, to the stock market~\cite{Park2007}. Lastly, unlike IIT the \eM{} analysis can be performed for data coming from a single channel. 

In this paper, we use the statistical complexity derived from an \eM{} analysis of neural activity to distinguish states of conscious arousal in fruit flies (\textit{D. melanogaster}). We analyse neural data collected from flies under different concentrations of isoflurane~\cite{CohenEneuro2016, CohenEneuro}. By analysing signals from individual electrodes and disregarding spatial correlations, we find that statistical complexity distinguishes between the two states of conscious arousal through temporal correlations alone. In particular, as the degree of temporal correlations increases, the difference in complexity between the wakeful and anaesthetised states becomes larger. In addition to measuring complexity, the \eM{} framework also allows us to assess the temporal irreversibility of a process- the difference in the statistical structure of the process when read forwards vs. backwards in time. This may be particularly important for wakeful brains which are thought to be sensitive to the statistical structure of the environment which runs forward in time~\cite{CohenEneuro, Hohwy2013, Tononi2010}. Using the nuanced characterisation of temporal information flow offered by the \eM{} framework~\cite{cryptCrutch}, we then analyse the time irreversibility and crypticity of the neural signals to further distinguish the conscious states. We find that the asymmetry in information structure between forward and reverse-time neural signals is reduced under anaesthesia.

The present approach singularly differentiates between highly random and highly complex information structure; accounts for temporal correlations beyond the Markov assumption; and quantifies temporal asymmetry of the process. None of the standard methods possesses all of these features within a single unified framework. Before presenting these results in detail in Sec.~\ref{sec:Complexity} and discussing their implications in Sec.~\ref{sec:discussion/conclusion}, we begin with a brief overview of the \eM{} framework we will use for our analysis.

\section{Theory: $\epsilon$-Machines and statistical complexity} \label{sec:Background}

To uncover the underlying statistical structure of neural activity that characterises a given conscious state, we treat the measured neural data, given by voltage fluctuations in time, as discrete time series. To analyse these time series, we use the mathematical tools of computational mechanics, which we outline in this section. We start with a general discussion on the ways to use time series data to infer a model of a system while placing \eMs{} in this context. Next, we explain how we construct \eMs{} in practice. Finally, we show how this can be used to extract a meaningful notion of statistical complexity of a process.

\subsection{From time series to $\epsilon$-Machines}
\label{sec:Bkg-eMs}

In abstract terms, a discrete-time series is a sequence of symbols $\mathbf{r} = (r_0, \ldots, r_{k}, \ldots)$ that appear over time, one after the other~\cite{Rabiner1989}. Each element of $\mathbf{r}$ corresponds to a symbol from a finite alphabet $\mathcal{A}$ observed at the discrete time step labelled by the subscript $k$. The occurrence of a symbol, at a given time step, is random in general and thus the process, which produces the time series, is stochastic~\cite{DoobStochastic}. However, the symbols may not appear in a completely independent manner, i.e., the probability of seeing a particular symbol may strongly depend on symbols observed in the past. These temporal correlations are often referred to as \textit{memory}, and they play an important role in constructing models that are able to predict the \textit{future} behaviour of a given stochastic process~\cite{Gu2012}.

Relative to an arbitrary time $k$, let us denote the future and the past partitions of the complete sequence as $\mathbf{r} = (\cev{r}, \vec{r})$, where the past and the future are $\cev{r} = (\ldots, r_{k-2},r_{k-1})$ and $\vec{r} = (r_{k}, r_{k+1}, \ldots)$ respectively. In general, for the prediction of the immediate future symbol $r_k$, knowledge of the past $\ell$ symbols $\cev{r}_{\ell} :=(r_{k-\ell}, \ldots, r_{k-2}, r_{k-1})$, may be necessary. The number of past symbols we need to account for in order to optimally predict the future sequence is called the Markov order~\cite{Gagniuc2017}.

In general, the difficulty of modelling a time series increases exponentially with its Markov order. However, not all distinct pasts lead to unique future probability distributions, leaving room for compression in the model. In a seminal work, Crutchfield and Young showed the existence of a class of models, which they called $\epsilon$-machines, that are provably the optimal predictive models for a non-Markovian process under the assumption of statistical stationarity~\cite{CrutchPRL1989, epsilonMachines2}. Constructing the $\epsilon$-machine is achieved by partitioning sets of \textit{partial} past observations $\cev{r}_{\ell}$ into \textit{causal states}. That is, two distinct sequences of partial past observations $\cev{r}_{\ell}$ and $\cev{r}_{\ell}'$ belong to the same causal state $S_i \in \mathcal{S}$, if the probability of observing a specific $\vec{r}$ given $\cev{r}_{\ell}$ or $\cev{r}_{\ell}'$ is the \textit{same}; that is 
\begin{gather}
     \cev{r}_{\ell} \sim_\epsilon \cev{r}_{\ell}' \quad\text{if}\quad P(\vec{r} \;|\; \cev{r}_{\ell}) = P(\vec{r} \;|\; \cev{r}_{\ell}'),
     \label{eq:equivRelation}
\end{gather}
where $\sim_\epsilon$ indicates that two histories correspond to the same causal state. The conditional probability distributions in Eq.~\eqref{eq:equivRelation} may always be estimated from a finite set of statistically stationary data via the naive maximum likelihood estimate, given by $P(r_k|\cev{r}_{\ell}) =\nu(r_k,\cev{r}_{\ell})/\nu(\cev{r}_{\ell})$, where $\nu(X)$ is the frequency of occurrence of sub-sequence $X$ in the data. For the case of non-stationary data, the probabilities obtained by this method will produce a non-minimal model that corresponds to a time-averaged representation of the time series. We now discuss how to practically construct an \eM{} for a given time series.

\subsection{Constructing \eMs{} with the CSSR algorithm}
\label{sec:Bkg-cssr}

Several algorithms have been developed to construct \eMs{} from time series data~\cite{Tino2001, CrutchPRL1989, Crutchfield1990}. Here, we briefly explain the \textit{Causal State Splitting Reconstruction} \textbf{(CSSR)} algorithm~\cite{CSSR2}, which we use in this work to infer \eMs{} predicting the statistics of neural data we provide as input. 

The CSSR algorithm proceeds to iteratively construct sets of causal states accounting for longer and longer sub-sequences of symbols. In each iteration, the algorithm first estimates the probabilities $P(r_k|\cev{r}_{\ell})$ of observing a symbol conditional on each length $\ell$ prior sequence and compares them with the distribution $P(r_k | \mathcal{S} = S_i)$ it would expect from the causal states it has so far reconstructed. If $P(r_k|\cev{r}_{\ell}) = P(r_k | \mathcal{S} = S_i)$ for some causal state, then $\cev{r}_{\ell}$ is identified with it. If the probability is found to be different for all existing $S_i$, then a new causal state is created to accommodate the sub-sequence. By constructing new causal states only as necessary, the algorithm guarantees a minimal model that describes the non-Markovian behaviour of the data (up to a given memory length), and hence the corresponding \eM{} of the process. 

The CSSR algorithm compares probability distributions via the \emph{Kolmogorov-Smirnov} \textbf{(KS)} test~\cite{Massey1951, Hollander2013}. The hypothesis that $P(r_k|\cev{r}_{\ell})$ and $P(r_k | \mathcal{S} = S_i)$ are identical up to statistical fluctuations is rejected by the KS test at the significance level $\sigma$ when a distance $\mathcal{D}_{KS}$~\footnote{The distance $\mathcal{D}_{KS} = \max | F(r_k | \mathcal{S} = S_i) - F(r_k | \unexpanded{\cev{r}}_{\ell})|$, where $F(r_k | \mathcal{S} = S_i)$ and $F(r_k | \unexpanded{\cev{r}}_{\ell})$ are cumulative distributions of $P(r_k | \mathcal{S} = S_i)$ and $P(r_k | \unexpanded{\cev{r}}_{\ell})$ respectively.} is greater than tabulated critical values of $\sigma$~\cite{Miller1956}. In other words, $\sigma$ sets a limit on the accuracy of the history grouping by parametrising the probability that an observed history $\cev{r}_{\ell}$ belonging to a causal state $S_i$, is mistakenly split off and placed in a new causal state $S_j$. Our analysis, in agreement with Ref.~\cite{CSSR2}, found that the choice of this value does not affect the outcome of CSSR within the tested range of $0.001 < \sigma < 0.01$. As a result, we set $\sigma = 0.005$.

As it progresses, the CSSR algorithm compares future probabilities for longer sub-sequences, up to a maximum past history length of $\lambda$, which is the only important parameter that must be selected prior to running CSSR in addition to $\sigma$. If the considered time series is generated by a stochastic process of Markov order $\ell$, choosing $\lambda < \ell$ results in poor prediction because the inferred $\eM{}$ cannot capture the long-memory structures present in the data. Despite this, the CSSR algorithm will still produce an \eM{} that is consistent with the approximate future statistics of the process up to order-$\lambda$ correlations~\cite{CSSR2}. Given sufficient data, choosing $\lambda \geq \ell$ guarantees convergence on the true $\eM{}$. One important caveat to note is that the time complexity of the algorithm scales asymptotically as $\mathcal{O} (|\mathcal{A}|^{2\lambda+1})$, putting an upper limit to the longest history length that is computationally feasible to use. 
Furthermore, the finite length of the time series data implies an upper limit on an `acceptable' value of $\lambda$. Estimating $P(r_k | \cev{r}_{\lambda})$ requires sampling strings of length $\lambda$ from the finite data sequence. Since the number of such strings grows exponentially with $\lambda$, a value of $\lambda$ that is too long relative to the size $N$ of the data, will result in a severely under-sampled estimation of the distribution. A distribution $P(r_k | \cev{r}_{\lambda})$ that has been estimated from an under-sampled space is almost always never equal to $P(r_k | \mathcal{S} = S_i)$, resulting in the algorithm creating a new causal state for every string of length $\lambda$ it encounters. A bound for the largest permissible history length is $L(N) \geq \log_2 N/\log_2 |\mathcal{A}|$, where $L(N)$ denotes maximum length for a given data size of $N$~\cite{MartonAoP, CoverThomas}. Once these considerations have been taken into account, the \eM{} produced by the algorithm provides us with a meaningful quantifier of the complexity of the process generating the time series, as we now discuss.
 
\subsection{Measuring the complexity and asymmetry of a process}
\begin{figure*}[th]
    \centering
    \includegraphics[width=\textwidth]
    {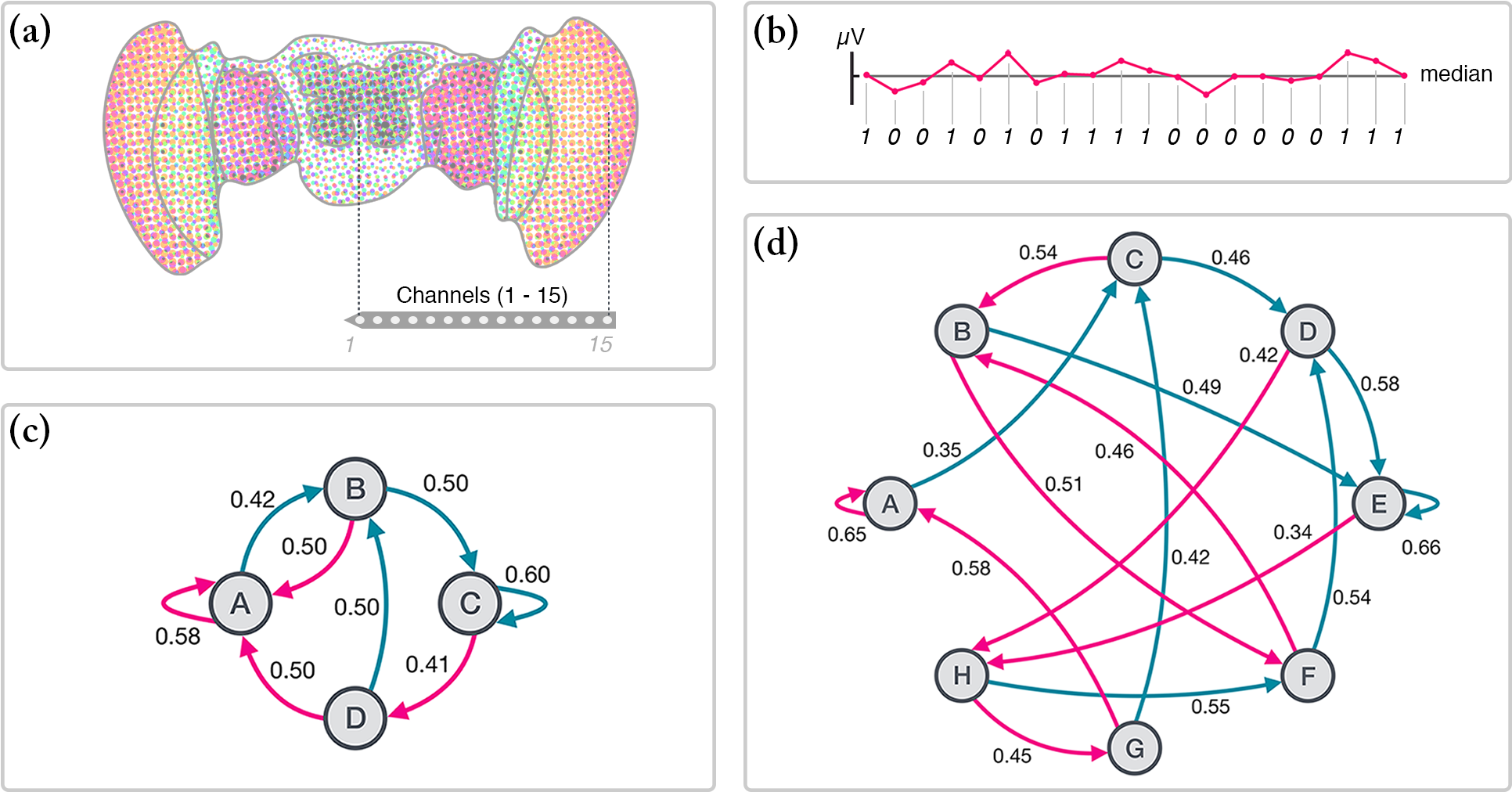}
    \caption{Evolution of experimental data from neural signals to \eMs.
\textbf{(a)} Representative schematic of \textit{D. melanogaster} brain (modified from Ref.~\cite{Paulk2013b}) depicted with probe and approximate channel locations. Each channel $c\in [1,15]$  samples around a localised region in the brain, with numerical labels ordered from the central ($c=1$) to peripheral ($c=15$) regions. 
\textbf{(b)} Example reading of a processed local field potential \textbf{(LFP)} for a single channel. Points along the x-axis represent LFP measurements at each sampling time step. The median LFP measurement of the sample is shown as the grey line bisecting data. LFP binarisation is determined via splitting over the median with the encoding scheme $0:$ LFP $\leq$ Median, and $1:$ otherwise. The \eMs{} are inferred by using the binary string as the input to the CSSR algorithm. 
\textbf{(c)} Digraph representation of the CSSR-inferred \eM{} for channel 1 readings of fly 1 under anaesthesia ($0.6$ vol.\% isoflurane) with $\sigma = 0.005$ and $\lambda = 3$. Graph vertices correspond to causal states. Vertex labels distinguishing causal states are assigned arbitrarily and do not imply state equivalence across multiple graphs. Directed edges correspond to transitions between causal states. Edge labels denote the probability (2 significant figures) of a transition occurring, and edge colour encodes the emitted symbol upon making the transition (1: Red, 0: Blue). The histories stored in the causal states for this \eM{} are visualised in Fig.~\ref{fig:em-with-histories}.
\textbf{(d)} Digraph representation of \eM{} for the wakeful ($0$ vol.\% isoflurane) level of conscious arousal for the same channel, fly, $\sigma$, and $\lambda$ as in (c). We report the forward-time statistical complexities $C_{\mu}^{a} = 1.88$ and $C_{\mu}^{w} = 2.96$ for (c) and (d) respectively.}
    \label{fig:workflow}
\end{figure*}

The output of the CSSR algorithm is the set of causal states and rules for transitioning from one state to another. That is, CSSR gives a Markov chain represented by a digraph~\cite{CrutchPRL1989, Gagniuc2017} $G(V,E)$ consisting of a set of vertices $v_i \in V$ and directed edges $\{i,j\} \in E$, e.g. Figs.~\ref{fig:workflow}(c) and (d). Using these rules, one can find $P(S_i)$, which represents the probability that the \eM{} is in the causal state $S_i$ at a any time. The Shannon entropy of this distribution quantifies the minimal number of bits of information required to optimally predict the future process; this measure, first introduced in Ref.~\cite{CrutchPRL1989}, is called the \emph{statistical complexity}:
\begin{gather}
    C_{\mu} := H\left[\mathcal{S}\right] = -\sum_i P(S_i) \log P(S_i).
     \label{eq:statComplexity}
\end{gather}

Formally, the causal states of a time series depend upon the direction in which the data is read~\cite{cryptCrutch}. The main consequence of this result is that the set of causal states obtained by reading the time series in the forward direction $\mathcal{S}^+$, are not necessarily the same as those obtained by reading the time series in the reverse direction $\mathcal{S}^-$. Naturally, this corresponds to potential differences in forward and reverse-time processes and the associated complexities, which is known as \textit{causal irreversibility}
\begin{gather}
    \Xi := C_{\mu}^{+} - C_{\mu}^{-},
    \label{eq:CausalIrrev}
\end{gather}
capturing the time-asymmetry of the process. 

Another (stronger) measure of time-asymmetry is \textit{crypticity}:
\begin{gather}
    d := 2C_{\mu}^{\pm}-C_{\mu}^{+}-C_{\mu}^{-}.
    \label{eq:crypticity}
\end{gather}
This quantity measures the amount of information hidden in the forwards and reverse \eMs{} that is not revealed in the future or past time series, respectively. Specifically, it combines the information that must be supplemented to determine the forwards \eM{} given the reverse \eMs{} and the information to determine reverse \eMs{} given the forwards \eM{}. In each case, this is equivalent to the difference between the complexity of a \emph{bidirectional} \eM{}, denoted $C_{\mu}^{\pm}$~\cite{cryptCrutch}, and that of the corresponding unidirectional machine. Throughout this manuscript, we implicitly refer to the usual forward-time statistical complexity $C_{\mu}^+$ when writing $C_{\mu}$, unless otherwise stated.

Finally, an operational measure for time-asymmetry is defined by the \textit{microscopic irreversibility}, which quantifies how statistically distinguishable the forwards and reverse \eMs{} are, in terms of the sequences of symbols they produce. If the forward-time \eM{} produces the same sequences with similar probabilities to the reverse-time \eM{}, then the process is reversible. Should a sequence available to $M^+$ be impossible for $M^-$ to produce, then the process is strictly irreversible.  Here, we assess the distinguishability between two \eMs{} by estimating the asymptotic rate of (symmetric) \emph{Kullback-Leibler} (\textbf{KL}) \emph{divergence} $\mathcal{D}_{KLS}$ between long output sequences; this measure is commonly applied to stochastic models~\cite{Yang2019}. Specifically
\begin{gather}
    \mathcal{D}_{KLS} = \mathcal{D}_{KL}(M^+ \| M^-) + \mathcal{D}_{KL}(M^- \| M^+),
\end{gather}
where $\mathcal{D}_{KL}$ is the regular, non-symmetric estimated KL divergence rate~\cite{Rached2004}. The KL divergence can be proved to be a unique measure that satisfies all of the theoretical requirements of information-geometry~\cite{amari2016book, Oizumi2016PNAS, amari2018}.

A few remarks are in order: in general, any one of the above measures vanishing does not imply that the other measures must also vanish. For instance, consider the case where the \emph{structures} of the forward ($M^+$) and reverse-time ($M^-$) \eMs{} are different but they happen to have the same complexities, i.e., $C_{\mu}^{+} = C_{\mu}^{-}$. Then, clearly we have $\Xi = 0$ but $d \ne 0$ and $\mathcal{D}_{KLS} \ne 0$. On the other hand, consider the case when $M^+$ and $M^-$ are the same; here, we have $\Xi = \mathcal{D}_{KLS} = 0$, yet may not have $d = 0$. This means that vanishing $\mathcal{D}_{KLS}$ implies that $\Xi=0$ (but not the converse, and not $d=0$). This turns out be an interesting extremal case because, while the forward and reverse processes are identical, the non-vanishing crypticity accounts for the information required to synchronise the corresponding \eMs{}, i.e., producing the joint statistics of the paired \eMs. Moreover, we can conclude that microscopic irreversibility is a stronger measure than causal irreversibility; this comes at the expense of computational cost, i.e., the former is harder to compute than the latter. In essence, each measure above represents a different notion of temporal asymmetry, with its own operational significance. Causal irrversibility and crypticity are information-theoretic constructs, while microscopic irrversibility is a information-geometric construct.

In the next section, we describe the experimental and analytical methods, as well as the results: that the statistical complexity and temporal asymmetry of the neural time series, taken from fruit flies, significantly differ between states of conscious arousal.

\section{Experimental results and analysis}
\label{sec:Complexity}

\subsection{Methods}
\label{sec:Methods}

We analysed local field potential \textbf{(LFP)} data from the brains of awake and isoflurane-anaesthetised \emph{D. melanogaster} (Canton S wild type) flies. Here, we briefly provide the essential experimental outline that is necessary to understand this paper. The full details of the experiment are presented in Refs.~\cite{CohenEneuro, CohenEneuro2016}. LFPs were recorded by inserting a linear silicon probe (Neuronexus 3mm-25-177) with 16 electrodes separated by 25 $\mu$m. The probe covered approximately half of the fly brain and recorded neural activity as illustrated in Fig.~ \ref{fig:workflow}(a). A tungsten wire inserted into the thorax acted as the reference. The LFPs at each electrode were recorded for 18s while the fly was awake and 18s more after the fly was anaesthetised (isoflurane, 0.6\% by volume, through an evaporator). Flies' unresponsiveness during anaesthesia was confirmed by the absence of behavioural responses to a series of air puffs, and recovery was also confirmed after isoflurane gas was turned off~\cite{CohenEneuro2016}.  

We used data sampled at 1kHz for the analysis~\cite{CohenEneuro2016}, and to obtain an estimate of local neural activity, the 16 electrodes were re-referenced by subtracting adjacent signals giving 15 channels which we parametrise as $c \in [1,15]$. Line noise was removed from the recordings, followed by linear de-trending and removing the mean. The resulting data is a fluctuating voltage signal, which is time-binned (1ms bins) and binarised by splitting over the median, leading to a time series, see Fig.~\ref{fig:workflow}(b). 

For each of the 13 flies in our data set, we considered 30 time series of length $N = 18,000$. These correspond to the 15 channels, labelled numerically from the central to peripheral region as depicted in Fig.~\ref{fig:workflow}(a), and the two states of conscious arousal. Using the CSSR algorithm~\cite{CSSR2}, we constructed \eMs{} for each of these time series as a function of  maximum memory length within the range $\lambda \in [2,11]$, measured in milliseconds. This is below the memory length $L(N) \sim 14$ beyond which we would be unable to reliably determine transition probabilities for a sequence of length $N$ (see Sec.~\ref{sec:Bkg-cssr})~\footnote{$L(N) \sim 14$ only serves as a lower bound on $\lambda$, past which CSSR is guaranteed to return incorrect causal states for the neural data. In practice, this may occur at even lower memory lengths than this limit. We observed this effect marked by an exponential increase in the number of inferred causal states for $\lambda > 11$, and thus excluded these memory lengths from the study.}. For a given time direction $\xi \in \{+:\text{forward},-:\text{reverse}\}$, we recorded the resulting $3,900$ \eM{} structures and their corresponding statistical complexities $C_{\mu}^{(\xi,\psi)}$, and grouped them according to their respective level of conscious arousal, $\psi \in \{w,a\}$ for awake and anaesthesia, channel location, $c$, and maximum memory length, $\lambda$. Thus, the statistical complexity we computed in a given time direction is a function of the set of parameters $\{\psi,c,\lambda\}$ for each fly, $f$. We also determined the irreversibility $\Xi$, crypticity $d$, and symmetric KL divergence rate $\mathcal{D}_{KLS}$ for each fly and again grouped them over the same set of parameters $\{\psi,c,\lambda\}$. While we found that not all the data is strictly stationary, in that the moving means of the LFP signals were not normally distributed, the conclusions we draw from them are still broadly valid. As mentioned in Sec.~\ref{sec:Bkg-eMs}, \eMs{} reconstructed from approximately stationary data are time-averaged models, and are likely to \emph{underestimate} the true statistical complexity of the corresponding neural processes.

We are principally interested in the differences the informational quantities $Q^\psi \in \{C_{\mu}^{(\xi,\psi)},\; \Xi^{\psi},\; d^{\psi},\; \mathcal{D}^{\psi}_{KLS}\}$ have over states of conscious arousal and thus consider
\begin{gather}
    \Delta Q := Q^{w} - Q^{a}, \label{eq:Dcmu}
\end{gather}
for fixed values of $\{f, c, \lambda\}$. Positive values of $\Delta Q$ indicate higher complexities observed in the wakeful state relative to the anaesthetised one. Finally, we use the notation $\langle Q^{\psi} \rangle_x$ to denote taking an average of any information quantity $Q^{\psi}$,  over a specific parameter $x \in \{f, c, \lambda\}$. For example $\langle \Delta C_{\mu}^+ \rangle_f$ means taking the fly-averaged difference in forward-time statistical complexity.

To assess the significance of each of the parameters $\psi$, $c$, $\lambda$, and $\xi$, or some combination of them, have on the response of the elements in the set $Q$ across flies, we conducted a statistical analysis using linear mixed effects modelling~\cite{Harrison2018lme} (\textbf{LME}). The LME analysis describes the response of a given quantity $\mathcal{Q}$ by modelling it as a multidimensional linear regression of the form
\begin{gather}
    \mathcal{Q} = \v{F}\gv{\beta} + \v{R}\v{b} + \mathcal{E}.
    \label{eq:glme}
\end{gather}
The resulting model in Eq.~\eqref{eq:glme} consists of a family of equations where $\mathcal{Q}$ is the vector allowing for different responses of a quantity $Q$ for each specific fly, channel location, level of conscious arousal, and time direction where applicable. Memory length $\lambda$, channel location $c$, state of conscious arousal $\psi$, and time direction $\xi$ (again, where applicable) are the parameters that $\mathcal{Q}$ responds to. To account for variations in the response caused by interactions between parameters (e.g. between memory length and channel location), we included them in the model. Letting $X = \{\lambda,c,\psi,\xi\}$ be the set of the parameters which may induce responses, we can write all the non-empty $k$-combinations between them as $\mathcal{F} = \{\lambda,c,\psi,\xi, \lambda c, \lambda\psi,...,\lambda c \psi \xi\} = \binom{X}{k}\backslash \emptyset$. The elements in $\mathcal{F}$ are known as the \emph{fixed effects} of Eq.~\eqref{eq:glme}, and are contained as elements within the matrix $\v{F}$. The vector $\gv{\beta}$, contains the regression coefficients describing the strength of each of the fixed effects $\mathcal{F}$.

In addition to fixed effects affecting the response of an element of $Q$ in our experiment, we also took into account any variation in response caused by known \emph{random effects}. In particular, we expected stronger response variations to be caused by correlations occurring between the channels within a single fly, compared to between channels across flies. These random effects are contained as elements of the matrix $\v{R}$, and the vector $\v{b}$ encodes the regression coefficients describing their strengths. Finally, the vector $\mathcal{E}$ describes the normally-distributed \emph{unknown} random effects in the model. The regression coefficients contained in the vectors $\gv{\beta}$ and $\v{b}$, were obtained via maximum likelihood estimation such that $\mathcal{E}$ are minimised. The explicit form of Eq.~\eqref{eq:glme} used in this analysis is detailed in the Appendix~\ref{app:lme}.

With the full linear mixed effects model given by Eq.~\eqref{eq:glme}, we tested the statistical significance of a fixed effect in $\mathcal{F}$. This was accomplished by comparing the log-likelihood of the full model with all fixed effects, to the log-likelihood of a reduced model with the effect of interest removed~\cite{Bates2015} (regression coefficients associated with the effect are removed). This comparison between the likelihood models is given by $\Lambda = 2 (h_{\text{full}} -h_{\text{reduced}})$, where $\Lambda$ is the likelihood ratio, $h_{\text{full}}$ is the log-likelihood of full model, and $h_{\text{reduced}}$ is the log-likelihood of the model with the effect of interest removed.

Under the null hypothesis, when a fixed effect does not have any influence on an informational quantity $Q$, i.e., the regression coefficients for the effect are vanishing, the likelihood ratio $\Lambda$ is $\chi^2$ distributed with degrees of freedom equal to the difference in the number of coefficients between the models. Therefore, we considered any fixed effect in the set $\mathcal{F}$ to have a statistically significant effect on a quantity if the probability of obtaining the likelihood ratio given the relevant $\chi^2$ distribution was less than 5\% ($p<0.05$). Thus, for each significant effect we report the fixed effect being tested, i.e., an element of $\mathcal{F}$, the obtained likelihood ratio $\chi^2(n-1)$ with $n$ associated degrees of freedom, and the associated probability $p$ of obtaining the statistic under the null hypothesis.

In addition to all the quantities in the set $Q^\psi$, the LME and likelihood ratio test was also performed for $\Delta Q$, in order to find the significant interaction effects of the parameters. Here, we also modelled $\Delta Q$ as dependent on a fixed effect in $\mathcal{F}$ as in Eq.~\eqref{eq:glme}, but excluding the parameter $\psi$ as it was already implicitly considered. Once the significant effects of memory length, level of conscious arousal, and channel location were characterised with our statistical analysis, we followed with post-hoc, paired $t$-tests for elements in $Q^{\psi}$ given by
\begin{gather}
    t = \frac{\langle \Delta Q\rangle_f}{s_f / \sqrt{|f|}},
    \label{eq:ttest-cmu}
\end{gather}
where $s_f$ is the standard deviation of $\langle \Delta Q \rangle_f$, and $|f| = 13$ is the sample size. The paired $t$-tests examine the nature of interactions between the parameters on a given quantity over the two states of conscious arousal. Positive $t$-scores indicate a quantity is larger for the wakeful state. We present the results of these analyses in the following sections, sorted categorically by whether time-direction is considered.

\subsection{Results}
\label{sec:results}

\subsubsection{Forward-time complexity results}
In order to observe the effects of isoflurane on neural complexity, we began by visually inspecting the structure of the reconstructed \eMs{} for the two levels of conscious arousal for the forward-time direction. We took special interest in observing the differences in the characteristics of the two groups of \eMs{} heralding the two levels of conscious arousal. Here, memory length $\lambda$ plays an important role. At a given $\lambda$, the maximum number of causal states that may be generated scales according to $|\mathcal{A}|^{\lambda}$~\cite{CSSR2}. In our case, the alphabet is binary, $\mathcal{A} = \{0,1\}$. This greatly restricts the space of \eM{} configurations available for short history lengths~\cite{Johnson2010}. For $\lambda = 2$ we can observe up to four distinct configurations, which is unlikely to reveal the difference based on conscious states. Given the previous findings~\cite{CohenEneuro}, we generally expected that the data from the wakeful state would present more complexity than those from the anaesthetised state. 

\begin{figure}[t]
    \centering
    \includegraphics[width=\columnwidth]{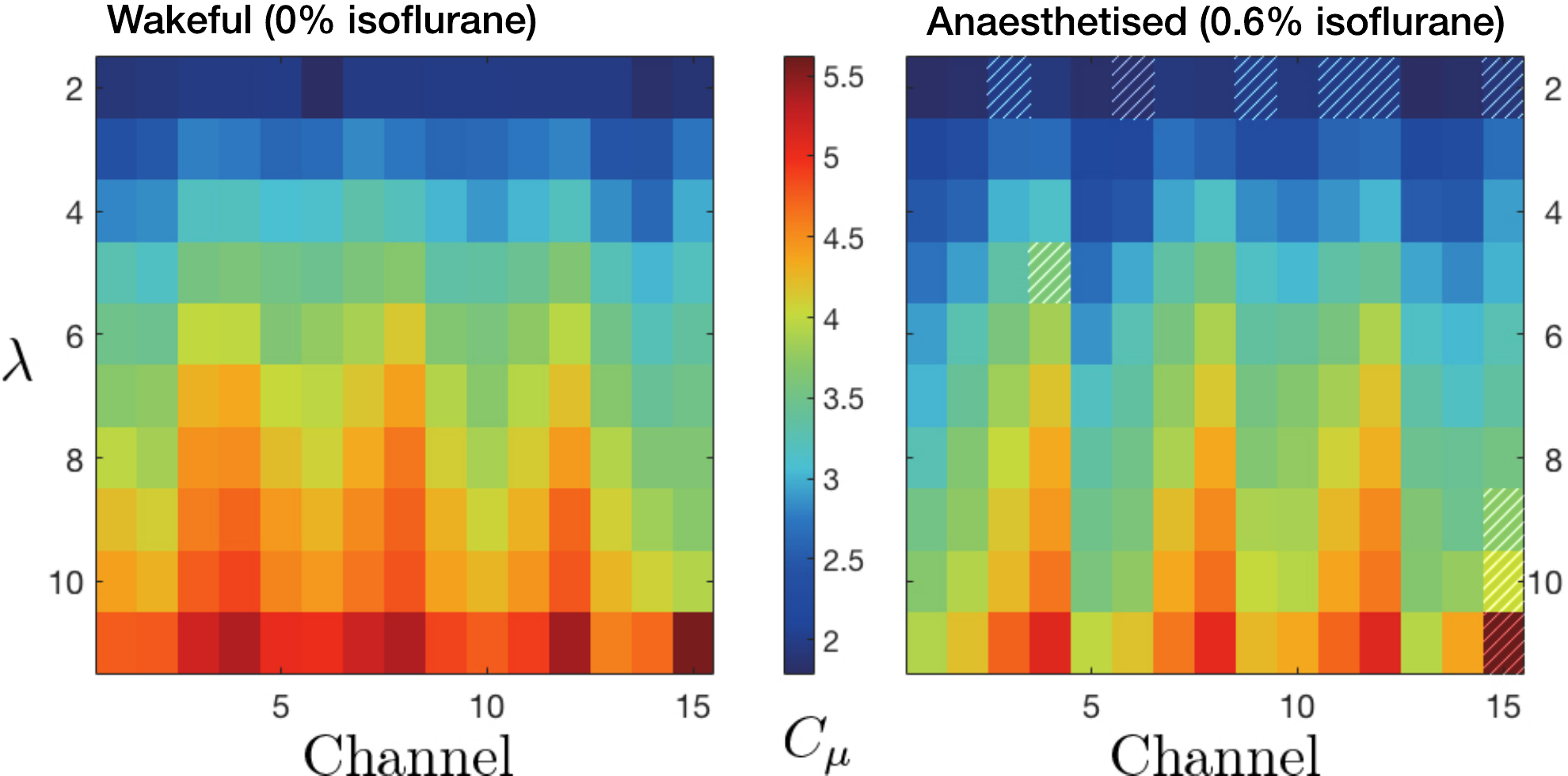}
    \caption{Colour map of statistical complexity response averaged over $(n=13)$ flies $\langle C^{(+,\psi)}_{\mu} \rangle_f$, during wakefulness (left) and anaesthesia (right), over channel location and memory length $\lambda$, measured in milliseconds. Hatched cells on the right sub-figure, show regions where $C_{\mu}$ did not decrease under anaesthesia.}
    \label{fig:Cmu-raw-response}
\end{figure}

Visual inspection of the directed graphs indeed suggested higher \eM{} complexity during the wakeful state compared to the anaesthetised state, at a given set of parameters $\{f,c,\lambda\}$. In particular, the data from the anaesthetised state tended to result in fewer causal states and overall reduced graph connectivity. Panels (c) and (d) of Fig.~\ref{fig:workflow} are examples of \eMs{} (channel 1 data recorded from fly 1, at maximum memory length $\lambda = 3$), where a simpler \eM{} is derived from the data under the anaesthetised condition. Differentiating between two conscious arousal states by visual inspection quickly becomes impractical because of the large number of  \eMs{}. Moreover, for large values of $\lambda$ the number of causal states is exponentially large and it becomes difficult to see the difference in two graphs. To overcome these challenges, we looked at a simpler index, the statistical complexity $C_\mu$, to differentiate between conscious arousal states. To systematically determine the relationships between $C_\mu$ and the set of variables $\{c,f,\psi\}$ we employed the LME analysis outlined in Sec.~\ref{sec:Methods}. We first tested whether $\lambda$ significantly affects $C_{\mu}$. We found $\lambda$ to indeed have a significant effect on $C_{\mu}$ ($\lambda$, ${\chi}^2(1)=443.64$, $p<10^{-16}$). Fig.~\ref{fig:Cmu-raw-response} shows that independent of the conscious arousal condition or channel location, $C_{\mu}$ increases with larger $\lambda$. This indicates that the Markov order of the neural data is much larger than the largest memory length ($\lambda=11$) we consider. Nevertheless, we have enough information to work with.

We then sought to confirm if the complexity of \eMs{} during anaesthesia are reduced, as suggested from visual inspection. Our statistical analysis indicates that $C_{\mu}$ is not invariably reduced during anaesthesia ($\psi$, ${\chi}^2(1)=0.212$, $p=0.645$) at all levels of $\lambda$ and all channel locations. This means that $C_\mu$ cannot simply indicate the causal arousal state without some additional information about time ($\lambda$) or space ($c$). In addition, we found that neither $c$ alone nor $c\psi$ strongly affects $C_\mu$. However, we found significant reductions in complexity when either the level of conscious arousal or the channel location, interacted with memory length ($\lambda\psi$, ${\chi}^2(1)=14.63$, $p=1.31\times10^{-4}$) and (${c\lambda}$, $\chi^2(14)=42.876$, $p=8.97\times 10^{-5}$) respectively. Moreover, the three-way interaction also had a strong effect ($\lambda\psi c$, ${\chi}^2(14)=24.00$, $p=0.0458$).

\begin{figure}[t]
\centering
\includegraphics[width=\columnwidth]{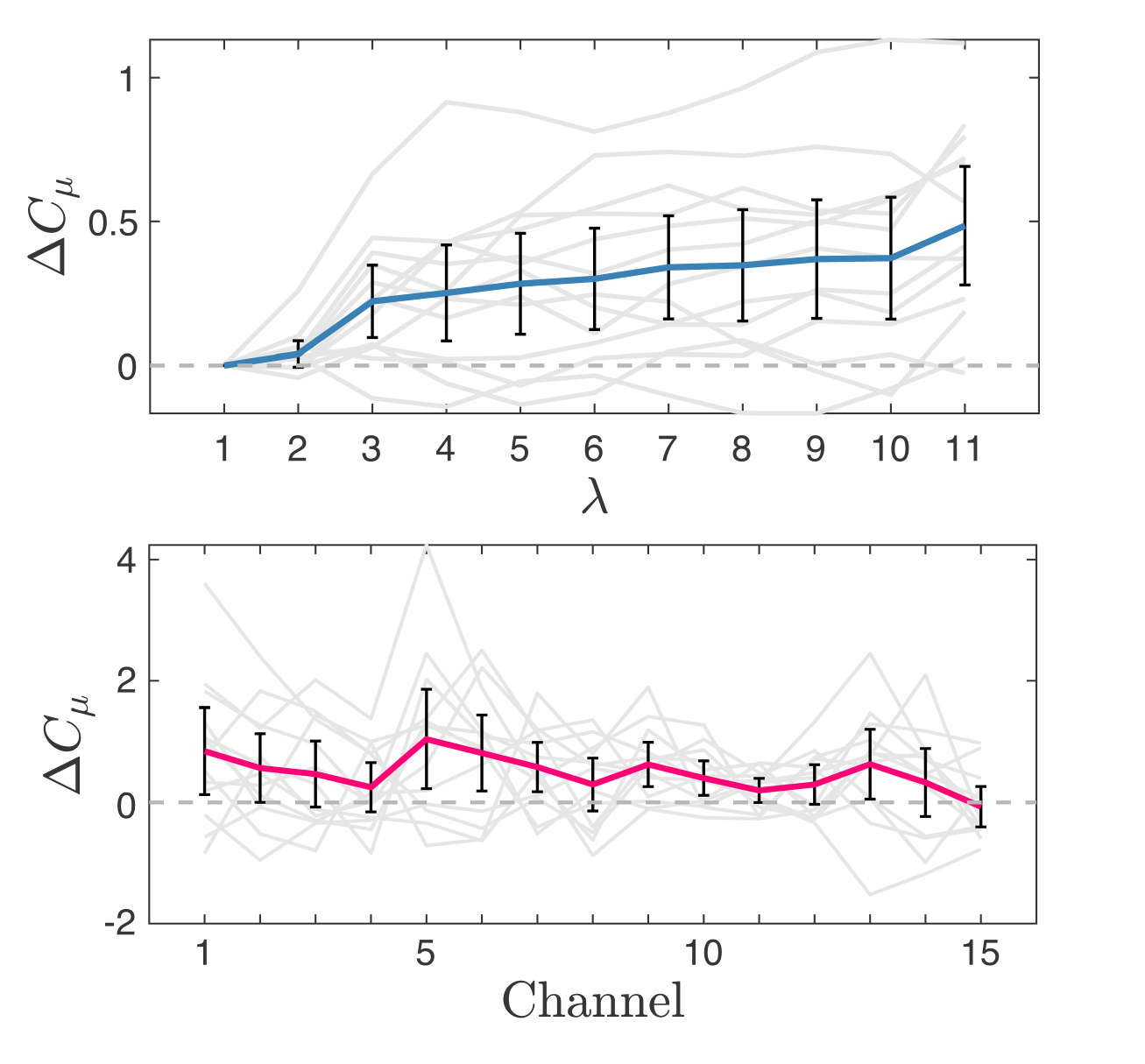}
\caption{Statistical complexity differences $\Delta C_{\mu} =C_{\mu}^{w} - C_{\mu}^{a}$ of \eMs{} between states of conscious arousal for: \textbf{(Top)} increasing memory length $\lambda$. Grey lines indicate complexity averages over channels per fly $(n=13)$, $\langle \Delta C_{\mu}\rangle_{c}$, while the blue line denotes the average over both channels and flies $\langle \Delta C_{\mu} \rangle_{c,f}$. Error bars are $95\%$ confidence intervals of the population. \textbf{(Bottom)} maximum memory length $\lambda = 11$ (in milliseconds), mapped throughout the fly brain (channels). Grey and red lines indicate the result per fly and the average over $(n=13)$ flies, $\langle \Delta C_{\mu} \rangle_{f}$, respectively. Error bars corresponding to the $95\%$ confidence intervals over the sample of files.}
\label{fig:DCmulti}
\end{figure}

As the three-way interaction between $\lambda$, $\psi$, and $c$ complicates interpretation of their effects, we performed a second LME analysis where we modelled \DCmu{} instead of $C_{\mu}$, thus accounting for $\psi$ implicitly. In doing so, we investigated whether the change in statistical complexity due to anaesthesia is affected by memory length $\lambda$ or channel location $c$. Using this model, we found a non-significant effect of $c$ on \DCmu{}, while a significant effect of $\lambda$ on \DCmu{} was seen ($\lambda$, ${\chi}^2(1)=20.97$, $p=4.65\times10^{-6}$), indicating that \DCmu{} overall changes with $\lambda$. Specifically, \DCmu{} tended to increase with larger $\lambda$ when ignoring channel location, as is evident in Fig.~\ref{fig:DCmulti} (top). Further, explaining our previous interaction between $\lambda$ and $\psi$, \DCmu{} was not clearly larger than $0$ for small memory length ($\lambda=2$; the top panel of Fig.~\ref{fig:DCmulti}). This suggests that the information to differentiate between states of conscious arousal is contained in higher order correlations. We also found that the interaction between ${\lambda}$ and channel location has a significant effect on \DCmu{} ($\lambda c$, ${\chi}^2(14)=37.19$, $p=6.90\times10^{-4}$), indicating that the effect of $\lambda$ is not constant across channels. Given that \DCmu{} overall increases with $\lambda$, we considered that that the largest \DCmu{} should occur at the largest $\lambda$. Fig.~\ref{fig:DCmulti} (bottom) examines \DCmu{} across channels at $\lambda=11$.

\begin{figure*}[t]
    \centering
    \includegraphics[width=\textwidth]
    {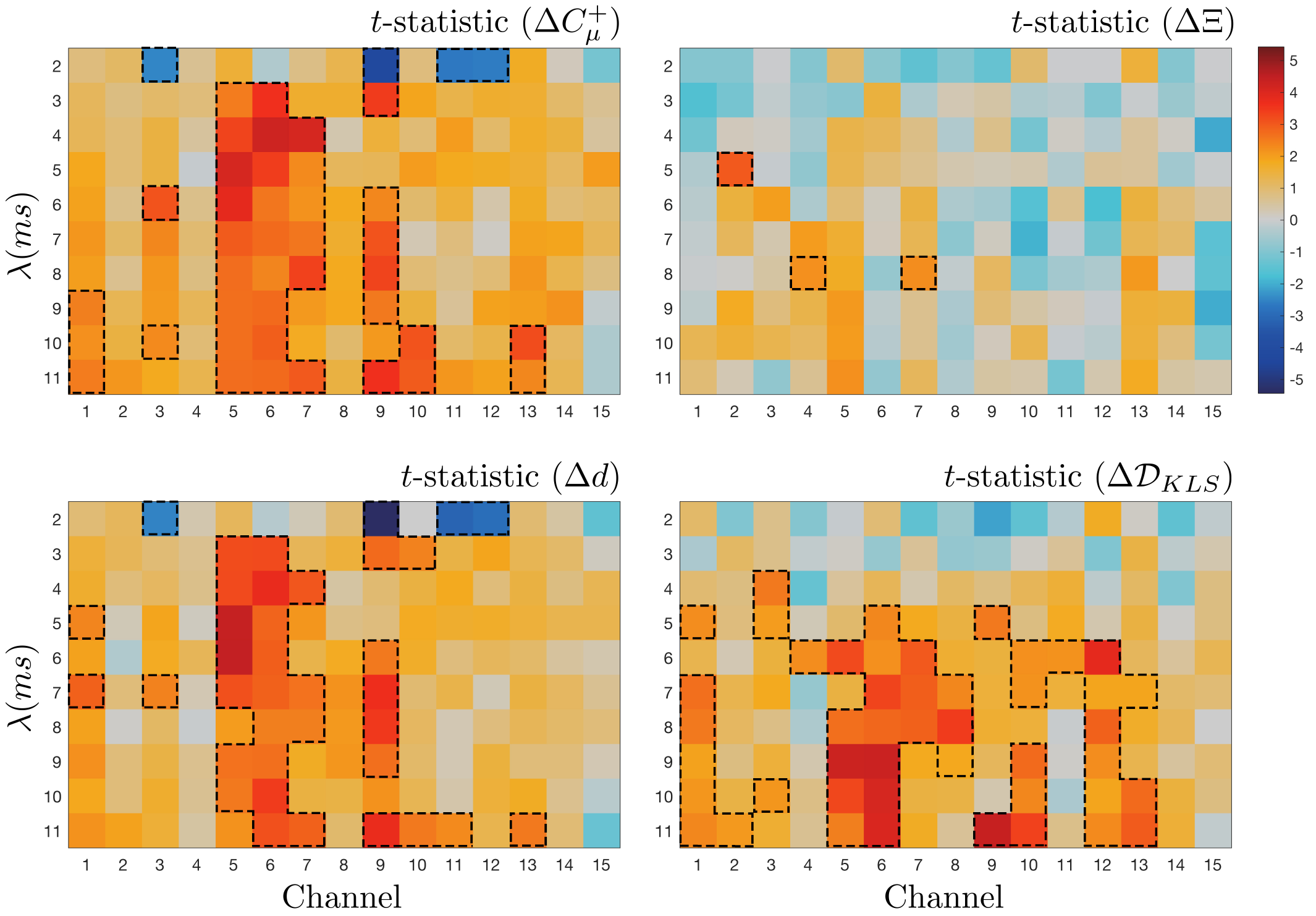}
    \caption{Colour map of two-tailed paired $t$-scores over channel location and memory length $\lambda$ for statistical complexity differences $\langle\Delta C^+_{\mu}\rangle_f = \langle C_{\mu}^{(+,w)} - C_{\mu}^{(+,a)}\rangle_f$ (top left); causal irreversibility differences $\langle \Delta \Xi \rangle_f = \langle \Xi^{w} - \Xi^{a} \rangle_f$ (top right); crypticity differences $\langle\Delta d\rangle_f = \langle d^{w} - d^{a}\rangle_f$ (bottom left); and the differences in KL divergence rate $\langle\Delta \mathcal{D}_{KLS}\rangle_f = \langle \mathcal{D}_{KLS}^{w} - \mathcal{D}_{KLS}^{a}\rangle_f$ (bottom right). The dotted lines indicate the memory length and channel locations that exceed $p < 0.05$ (uncorrected). The colour scale is consistent across all subplots.}
    \label{fig:heatmap}
\end{figure*}

To further break down the interaction between $\lambda$ and $c$, we performed a one sample $t$-test at each value of memory length and channel location to find regions in the parameter space $(\lambda, c)$ where $C_{\mu}$ reliably differentiates wakefulness from anaesthesia across flies. We plot the $t$-statistic at each parameter combination in the top-left panel of Fig.~\ref{fig:heatmap}, outlining regions in the parameter space where \DCmu{} is significantly greater than $0$ (with $p<0.05$, uncorrected, two-tailed), finding that the majority of the significance map is directed towards positive values of the $t$-statistic. However, only a subset of $(\lambda,c)$ cells contain values which are significantly different from $0$. Interestingly, we observed that for $\lambda=2$, \DCmu{} is actually significantly negative, corresponding to greater complexity during anaesthesia, not during wakefulness. This marks $\lambda=2$ as anomalous relative to other levels of $\lambda$, and this reversal of the direction of the effect of anaesthesia likely contributed to the interaction between $\lambda$ and $\psi$.

Disregarding $\lambda=2$, we find \DCmu{} to be significantly greater than $0$ for channels 1, 3, 5-7, 9, 10, and 13, at varying levels of $\lambda$. As expected from our reported interaction between $\lambda$ and $c$, we observe \DCmu{} to already be significantly greater than $0$ at small $\lambda$ for channels 5-7, while \DCmu{} only became significantly greater at larger $\lambda$ for channels 1, 3, 10 and 13. Further, other channels such as the most peripheral channel ($c=15$) did not have \DCmu{} significantly greater than $0$ at any $\lambda$. All significance results, due to LME tests, are reported in Table~\ref{tab:LMEresults}.

The above results suggest that the measured difference in complexity is present across various brain regions (top-left panel of Fig.~\ref{fig:heatmap}), and that it grows as longer temporal correlations are taken into account (up to the largest value $\lambda=11$  tested). While Fig.~\ref{fig:DCmulti} shows a continued increase in the difference of statistical complexity, $\Delta C_{\mu}$, as a function of $\lambda$, we did not analyse longer history lengths, due to limitations in the amount of the data and stability of the estimation of $C_{\mu}$. In addition to this general observation of increasing \DCmu{} with $\lambda$, we observe that, remarkably, some brain regions  discriminate the conscious arousal states with a history length of only 3. One trivial explanation for this effect is that under anaesthesia, the required memory length is indeed $\lambda =2$, while the optimal $\lambda$ for awake is much larger. However, a quick observation of Fig.~\ref{fig:Cmu-raw-response} rules out this simple possibility; under both wakeful and anaesthetised states, $C_{\mu}$ continues to increase.

It is likely, however, that the tested range for $\lambda$ remains below the Markov order of the neural data; this is clearly indicated by the lack of a plateau in statistical complexity in Fig.~\ref{fig:DCmulti}. This suggests that we are far from saturating the Markov order of the process, and with more data we would be able to further distinguish between the two states. Future analyses with longer time series would also contribute to our understanding of the Markov order (maximum memory length) differences between the two states of conscious arousal. Nevertheless, our results, in Figs.~\ref{fig:DCmulti} and~\ref{fig:heatmap}, demonstrate that saturation of Markov order is not required for discrimination between conscious arousal states. This finding has a practical implication about the empirical utility of \eMs{}; even if the history length is too low, the inferred \eM{} and its statistical complexity can be useful. We now discuss the temporal asymmetry of neural processes.

\subsubsection{Temporal asymmetry}\label{sec:cryp}

Unlike other complexity measures, we obtain a distinct \eM{} from each given time series, and for each direction we read the time series, i.e., forward or backward in time. Based on the notion that wakeful brains should be better at predicting the next sensory input~\cite{CohenEneuro}, we expect that anaesthesia should alter the information structures depending on the time direction. Our expectation translates to the following three hypotheses:
\begin{enumerate}
    \item Causal irreversibility ($\Xi := C_{\mu}^{+} - C_{\mu}^{-})$, which is purely based on the summary measure of statistical complexity, should be higher for awake but lower for anaesthetised brains;
    
    \item Crypticity ($d := 2C_{\mu}^{\pm}-C_{\mu}^{+}-C_{\mu}^{-}$) should be higher for wakeful than anaesthetised brains;

    \item Symmetric KL divergence rate ($\mathcal{D}_{KLS}$) should behave similarly.
\end{enumerate}

On visual inspection of the variation in $\Xi$ for the wakeful and anaesthetised conditions, both appeared close to zero,  suggesting that $\Xi$ would not have significant dependence on the condition. This impression was confirmed statistically with two-tailed $t$-tests against zero with corrections for multiple comparisons, as shown in the top-right panel of Fig.~\ref{fig:heatmap}. Thus, Hypothesis 1 above, that irreversibility should be higher for wakeful over anaesthetised brains, is not supported by the data. However, as mentioned earlier, vanishing $\Xi$ does not imply that either $d=0$ or $\mathcal{D}_{KLS}=0$. To rule out the possibility that the information structure of \eMs{} are different when read forwards, as opposed to backwards, depending on the condition, we also tested the latter two hypotheses.

With respect to crypticity, first, visual inspection of the two-tailed $t$-score map, which compares crypticity for the wakeful $d^{w}$ and anaesthetised $d^{a}$ conditions (bottom-left panel of Fig.~\ref{fig:heatmap}) strongly implies that crypticity is larger in the former compared to the latter. This difference is largest over channels 5-7 and 9. To systematically evaluate this impression, we used LME statistical analysis (described in Sec.~\ref{sec:Methods}) to determine the relationships between crypticity, $d$, and the set of variables $\{c,\lambda,\psi\}$ we employ. As expected, we found that both memory length ($\lambda$) and level of conscious arousal ($\psi$) significantly affected crypticity ($\lambda$, $\chi^2(1)=470.5$, $p<10^{-16}$) and ($\psi$, $\chi^2(1)=5.896$, $p=0.0152$) respectively. Crypticity also depended on a significant interaction between memory length and condition ($\lambda \psi$, $\chi^2(1)=6.119$, $p=0.0134$). Specific increases in crypticity around the middle brain region (bottom-left panel of Fig.~\ref{fig:heatmap}) were also evident, with a strong interaction between channel location and memory length ($\lambda c$, $\chi^2(14)=35.86$, $p=1.09\times10^{-3}$), which is similar to the result obtained for \DCmu{}. This LME analysis, together with the direction of effects in Fig.~\ref{fig:heatmap} (bottom-left) strongly confirms our Hypothesis 2.

Furthermore, as a more direct measure of microscopic structure, we also analysed the symmetric KL divergence rate, $\mathcal{D}_{KLS}$. Again, the two-tailed $t$-score map (Fig.~\ref{fig:heatmap}, bottom-right panel) showed support for our hypothesis. Our formal statistical analysis with LME confirmed a critical interaction between memory length and condition ($\lambda \psi$, $\chi^2(1) = 15.37$, $p<10^{-16}$), meaning that time-asymmetric information structure is lost due to anaesthesia, especially when a long memory length is taken into account. (We also note other significant effects: mainly the effect of memory length ($\lambda$, $\chi^2(1) = 127.4$, $p<10^{-16}$) and interaction between memory length and channel location ($\lambda c$, $\chi^2(14) =85.81$, $p<10^{-16}$). Again, all significant results, due to LME tests, are reported in Table~\ref{tab:LMEresults}.

\begin{figure*}[th]
    \centering
    \includegraphics[width=\textwidth]
    {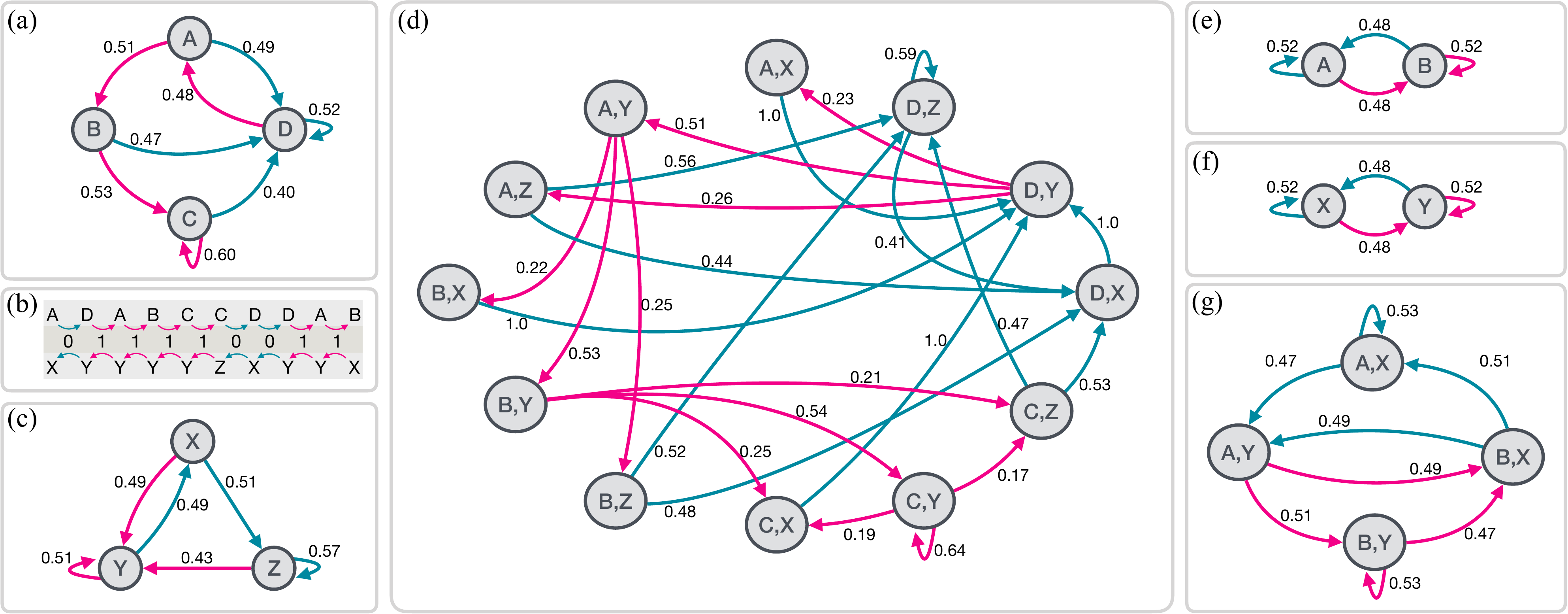}
\caption{Exemplary digraph representations of \eMs{} for wakeful  \textbf{(a-d)} and anaesthetised  \textbf{(e-g)} conditions for forward-time \textbf{(a, e)}, reverse-time \textbf{(c, f)}, and bidirectional \textbf{(d, g)} analyses, all constructed from channel 5 in fly 7, at memory length $\lambda=3$. Panel \textbf{(b)} gives an example emission sequence and causal state sequence for forward and reverse-time \eM{} pair (a) and (c). The vertex labelling denoting causal states in (a-d) is consistent to show composition of forward and reverse-time \eMs{} in the bidirectional \eM{}.
The \eMs{} for the wakeful condition have statistical complexity of $C_{\mu}^{(+,w)} = 1.76$, $C_{\mu}^{(-,w)} = 1.50$, and $C_{\mu}^{(\pm,w)}=3.25$. In this example the process is irreversible for all three quantities.
The \eMs{} for the anaesthetised condition have statistical complexity of $C_{\mu}^{(+,a)} = C_{\mu}^{(-,a)} = 1.0$ and $C_{\mu}^{(\pm,a)} = 1.9989$. The process is causally and microscopically reversible, but has finite crypticity.}\label{fig:fwdbwd}
\end{figure*}

Taken together, these results show that the relative complexity of the forward versus reverse direction, as measured by causal irreversibility, does not distinguish between the wakeful and anaesthetised states. However, our crypticity results demonstrate that, under anaesthesia, the structures of the forward and reverse processes are relatively similar, whereas during wakefulness their structures differ. Fig.~\ref{fig:fwdbwd} demonstrates this effect with exemplar \eMs{} reconstructed from a representative channel, from which we derived six distinct \eMs{}: three for wakeful (a, c, d), and three for anaesthetised (e-g) flies. Panel (b) shows how the time series and the transitions in the causal states of forward, reversed, and bidirectional \eMs{} are related.

Our finding, that causal irreversibilities were not above zero for wakeful brains, corresponds to the fact that complexities of forward and reverse \eMs{} were not significantly different. However, the bidirectional \eMs{} for the wakeful condition were substantially more complex than those for the anaesthetised condition. The statistical complexity of bidirectional \eMs{} should equal that of forward or reverse \eMs{} if the process is completely time symmetric and deterministic~\cite{cryptCrutch}, resulting in zero crypticity. However, for non-deterministic processes, additional information for synchronising the forward and reversed process may be needed, which would mean $d>0$. For instance, in Fig.~\ref{fig:fwdbwd}(e-f), if we are told that the forward machine is in causal state $A$, we need extra information to determine whether the reversed machine is in causal state $X$ or $Y$. Yet, the detailed structure of the forward and reverse machines are the same in this example. Our analysis is supplemented with a study of the symmetric KL divergence rate between forwards and backwards processes, which measures the distance between the reconstructed \eMs{}. In other words, crypticity and symmetric KL divergence rate quantify two different notions of temporal asymmetry; the former is information theoretic, and the latter is information-geometric. Indeed, in general we find that the processes in the two directions are different in both ways and, further, their difference varies significantly between conditions, as shown in the bottom two panels of  Fig.~\ref{fig:heatmap}.

\section{Discussion}
\label{sec:discussion/conclusion}

Discovering a reliable measure of conscious arousal in animals and humans remains one of the major outstanding challenges of neuroscience. The present study addresses this challenge by connecting a complexity measure to the degree of conscious arousal, taking a step forward to strengthening the link between physics, complexity science, and neuroscience. Here, we have taken tools from the former and have applied them to a problem in the latter. Namely, we have studied the statistical complexity and time asymmetry of neural recordings in the brains of flies over two states of conscious arousal: awake and anaesthetised. We have demonstrated that differences between these macroscopic states can be revealed by both the statistical complexity of local electrical fluctuations in various brain regions, and various measures of temporal asymmetry of hidden models that explain their behaviour. Specifically, we have analysed the single-channel signals from electrodes embedded in the brain using the \eM{} formalism, and quantified the statistical complexity $C_{\mu}$, causal and microscopic reversibility $\Xi$ \& $\mathcal{D}_{KLS}$, and crypticity $d$ of the recorded data for 15 channels in 13 flies over two states of conscious arousal. We find the statistical complexity is larger on average when a fly is awake than when the same fly is anaesthetised ($\Delta C_{\mu} > 0$; Figs.~\ref{fig:DCmulti} and~\ref{fig:heatmap}), and that the structural complexity of information and its time reversibility captured by crypticity and KL rate are also reduced under anaesthesia ($\Delta d > 0$ and $\Delta \mathcal{D}_{KLS} > 0$;~Fig.~\ref{fig:heatmap}).

As we have demonstrated in this study, the local information contained within a single channel can contain information about global conscious states, which are believed to arise from interactions among many neurons. Theoretically, single channels can reflect the complexity of the multiple channels due to the concept of Sugihara causality~\cite{Sugihara2012}. This arises due to any one region of the brain causally interacting with the rest of the brain, making the temporal correlation in a single channel time series contain information about the spatial correlations, i.e., information that would be contained in multiple channels. With this logic, Ref.~\cite{Tajima2015} infers the complexity of the multi-channel interactions from a single channel temporal structure of the time series. This is often known as the backflow of information in non-Markovian dynamics~\cite{BreuerPRL2009}. The periodic structure of statistical complexity observed across channels in Fig.~\ref{fig:Cmu-raw-response}, demonstrates an unexpected example of spatial effects present in our study -- one that was not observed with conventional LFP analyses. This observation may provide a motivation for multi-channel analyses.

While we already find differences between conscious states in the single channel based \eM{} analysis, it would be beneficial to extend the present analysis to the multi-channel scenario, in which \eM{} can be contrasted with the methods of IIT~\cite{Casali2013, Casarotto2016, Tononi2004, Tononi2016, Oizumi2014, Mediano2019, Barrett2011, Tegmark2016, Oizumi2016PNAS, Oizumi2016PLOS}. Formal comparison of the distinguishing power of conscious states among proposed methods (such as those in Ref.~\cite{Engemann2018,Sitt2014}) will contribute to refining models and theories of consciousness.

Our results can be informally compared with a previous study, where the \emph{power spectra} of the same data in the frequency domain~\cite{CohenEneuro} was analysed. There, a principal observation was the power in low-frequency signals in central and peripheral regions, which was more pronounced in the central region (corresponding to channel 1-6 in this study). Our \eM{} analysis here reveals that the region between periphery and centre (channels 5-7) shows most consistent difference in $C_{\mu}$ across history length $\lambda > 2$. Ultimately, the reason for this difference is due to our distinct approach, insofar as \eMs{} are provably the optimal predictive models of a large class of time series that take into account higher order correlations memory structure~\cite{CrutchPRL1989, epsilonMachines2}. Thus, our application of \eMs{} contrasts with the power spectra analysis, by considering these higher order correlations for very high-frequency signals, instead of only two-point correlations in both high- and low-frequency signals. Finally, the top-left panel of Fig.~\ref{fig:heatmap} shows that in regions corresponding to channels 1 and 13, the differences in the conditions are only seen at high values of $\lambda$.

\begin{figure}[t]
\centering
\includegraphics[width=\columnwidth]{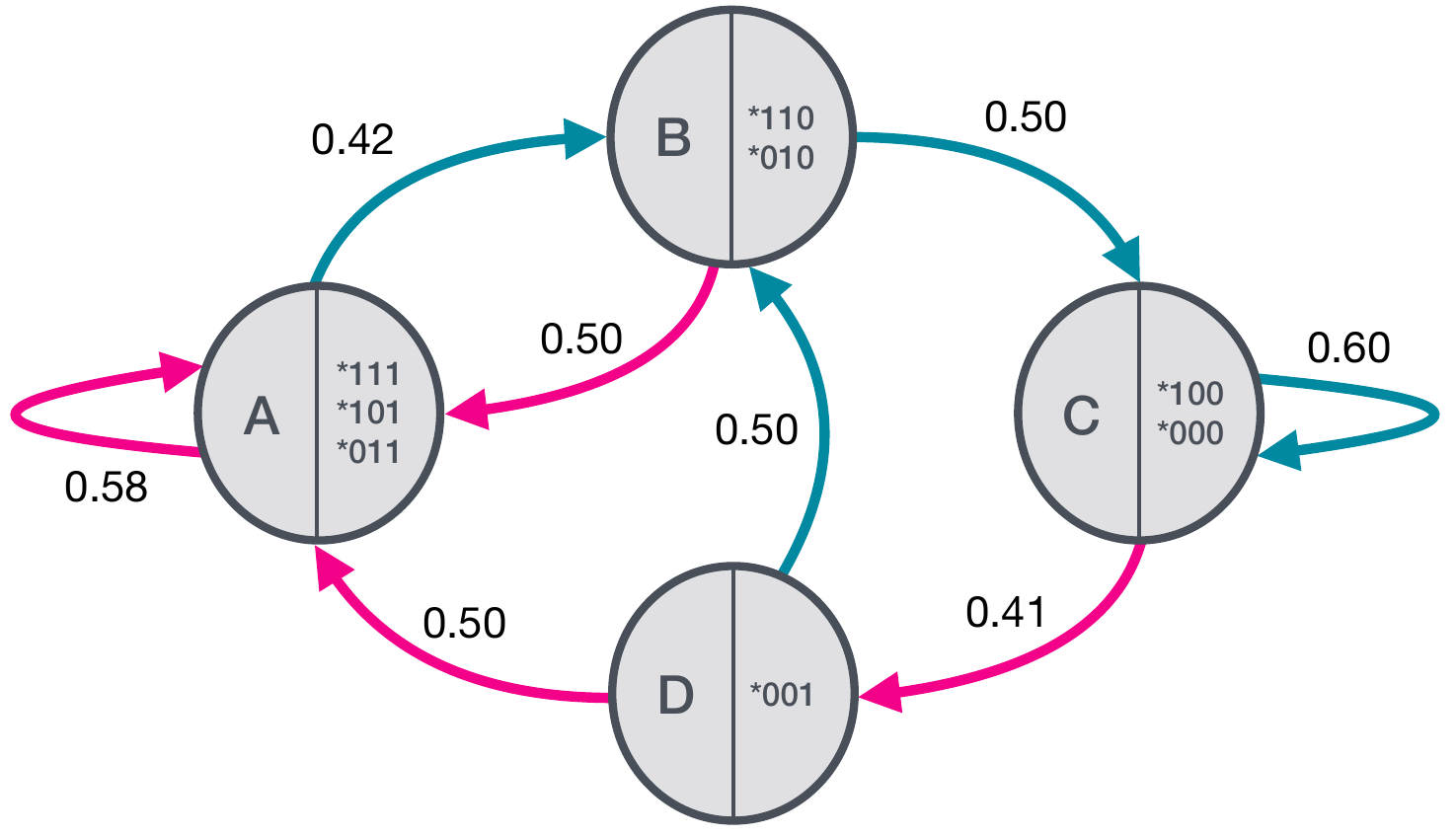}
\caption{\eM{} for same channel, fly, conscious state as Fig.~\ref{fig:workflow}(c), but with histories stored in each causal state explicitly stated. The sequences after the asterisk $*$ represent the sequence of symbol observations with the most recent observed symbol on the far right. Sequences collected within a causal state (grey circle) warrant significantly different future statistics to observed sequences in other causal states. The red lines emit a ``1" upon transition, and blue lines emit ``0"s.}
\label{fig:em-with-histories}
\end{figure}

Our multi-time analysis further reveals an interesting effect when we look more closely at, e.g., the anaesthetised \eM{} example shown in Fig.~\ref{fig:workflow}(c). When we examine the binary strings belonging to each causal state, we find a clear split between active (consecutive strings of ones) and inactive (consecutive strings of zeros) neural behaviour corresponding to the left and right hand sides of Fig.~\ref{fig:em-with-histories} respectively. Previous studies have demonstrated an increase in low-frequency LFP and EEG power for mammals and birds during sleep and anaesthesia, mediated by similar neural states of activity and inactivity known as `up' and `down' states~\cite{Sarasso2015, Lewis2012}. A similar phenomenon has recently been observed in sleep deprived flies~\cite{raccuglia2019}. Consistent with other studies, our study, using general anaesthesia, does not observe this slow oscillations. Future studies with more formal comparisons between up and down states and \eMs{}, in both theory and computer simulations, may be a fruitful avenue for further research in this regard.

An analysis in terms of \eMs{} has also allowed us to discriminate between levels of conscious arousal by examining causal structures found in both forward and reverse time directions. Based on our previous finding~\cite{CohenEneuro} as well as related concepts in temporal predictive~\cite{Hohwy2013, Friston2010, Tani1999, Bastos2012} and causal matching~\cite{Tononi2010}, we hypothesised that the wakeful brain may be tuned to causal structures of the world, which run forward in time, and thus \eMs{} would be more complex for forward than reverse readings. Further, we hypothesised that such temporally tuned structural matching will be lost under anaesthesia. Our results (Sec.~\ref{sec:cryp}) are highly intriguing in three ways. First, near-zero \emph{causal} irreversibility indicates that reducing the structural complexity to a simple index is not enough to capture effects on the information structure that are sensitive to the direction of time. This is the case regardless of the level of consciousness (at least at the timescales of this study). Second, nonzero crypticity indicates that the underlying information structure is not symmetric in time. More precisely, the signals themselves encode different amounts of information when run forwards as opposed to backwards. Third, the KL divergence rate analysis definitively demonstrated the existence of greater temporal irreversibility in the wakeful as opposed to the anaesthetised state. Having said this, we are limited in drawing strong conclusions due, in part, to the relatively small observed effect size of $\Xi$, likely a consequence of our relatively small data set. Despite this, even at millisecond time scales, our study successfully identifies significant differences in the time direction of the neural recordings.

Identifying the decrease in temporal-reversibility due to anaesthesia in tandem with complexity is of broad interest in neuroscience. While some physicists and neuroscientists have conjectured links among physics, the brain, and even consciousness through the lens of the direction of the time, their accounts have remained rather speculative, and not built on any solid theoretical foundations (for related and  alternative theoretical foundations, see the work by Cofr\'{e} and colleagues~\cite{cofre2019,cofre2018}). For example, using reversely played movies, the sensitivity to the direction of time is shown to differ across brain regions in humans~\cite{hasson2008}. In animal studies, some populations of neurons (in the hippocampus) are known to become activated in a particular sequential order while the animal experiences a particular event. For example, in anticipation of the event, the neurons activate in a forward direction, but in retrospection, the neurons activate in reverse order~\cite{Diba2007}. While direct links between these empirical findings and the \eM{} framework remains elusive, we foresee that our unified theoretical and analytical framework can potentially bridge this gap in the future.

Our study is not the first to apply complexity measure in consciousness research. Indeed, many definitions and measures of complexity have been proposed in the literature (see Ref.~\cite{Edmonds1997} for a list). Moreover, there is a flow of ideas going the other way as well~\cite{PhysRevLett.119.225301, PhysRevA.97.052320, QIIT}. However, many, if not most, of these measures cannot account for temporal correlations (memory), temporal asymmetry, or differentiate between random and structured processes. Our interdisciplinary study, based on \eMs{}, opens up new possibilities; physics can improve its theoretical constructs through the application of tools to empirical data, while neuroscience can benefit from rigorous quantitative tools that have proven their physical basis across different spatio-temporal scales. Among those complexity measures, $C_{\mu}$ can be easily interpreted in terms of temporal structure~\cite{whyCmu}, as it has a direct relation to process predictability and memory requirements. We emphasise that statistical complexity $C_{\mu}$ derived from \eMs{}, drastically differentiates itself from other scalar complexity indices such as Lempel-Ziv complexity~\cite{schartner2015}. For one Lempel-Ziv complexity is maximal for a random noise process whereas statistical complexity for the same process is zero (see Eq.~\eqref{eq:statComplexity}). In addition, the notion of temporal reversibility available in the \eM{} framework has no counterpart in Lempel-Ziv complexity. This is a critical difference since it is known that a low-complexity forward-time \eM{} consisting of only two causal states can have a very high-complexity reverse-time \eM{} with countably infinite states~\cite{Ellison2011}. Thus, explicitly considering the influence of time is critical for addressing questions about complexity. When coupled with our results, we can conclude that anaesthetised brains become less structured, more random, more reversible, and approach a stochastic process with a smaller memory capacity compared to the wakeful brains.

Overall, our results suggest that measures of complexity extracted from \eMs{} might be able to identify further structures that are affected by anaesthesia at different spatial and temporal scales. It is also likely that applying a similar analysis to other data sets, in particular, human EEG data will lead to new discoveries regarding the relationship between consciousness and complexity that can be retrieved simply at the single channel level.

\begin{acknowledgements}
RNM, FAP, NT, KM acknowledge support from Monash University's Network of Excellence scheme and the Foundational Questions Institute (FQXi) grant on \textit{Agency in the Physical World}. AZ was supported through Monash University's Science-Medicine Interdisciplinary Research grant. DC was funded by an Overseas JSPS Postdoctoral Fellowship. NT was funded by Australian Research Council Discovery Project grants (DP180104128, DP180100396). NT and CD were supported by a grant (TWCF0199) from Templeton World Charity Foundation, Inc. We thank Felix Binder, Alec Boyd, Mile Gu, Rhiannon Jeans, and Jayne Thompson for valuable comments. KM is
supported through Australian Research Council Future Fellowship FT160100073.
\end{acknowledgements}

\section*{Appendix}

\subsection{Linear mixed-effects model}
\label{app:lme}
In this section, we demonstrate an example of an LME analysis for the case of statistical complexity $C_{\mu}$ in the forward time direction. For the case when time direction $\xi$ is included as an effect, the only change this makes to the process is increasing the dimensions of the effects matrix $\mathbf{F}$. Performing an LME analysis on other quantities used in this study like crypticity or KL rate follow the same procedure outlined here.

The main goal of the LME analysis we perform in this study is to determine the degree of contributions each and combinations of memory length ($\lambda$), channel location ($c$), and level of conscious arousal ($\psi$) have on statistical complexity $C_{\mu}$. LME accomplishes this by modelling statistical complexity as a general linear regression equation (Eq.~\eqref{eq:glme}), whose response is predicted by the aforementioned parameters $\lambda$, $c$, and $\psi$. In this Appendix, we show the exact form of the linear regression equation used in this analysis, while referring to the terminology introduced in the methods (Sec.~\ref{sec:Methods}).

We begin by restating Eq.~\eqref{eq:glme} for the case of statistical complexity, $\mathcal{C} = \v{F}\gv{\beta} + \v{R}\v{b} + \mathcal{E}$, which has the form of a general multidimensional linear equation. We will set aside the right hand side of the equality for now. On the left hand side, statistical complexity takes the form of a column vector $\mathcal{C}$. Each row corresponds to the unique response of $C_{\mu}$, at a specific selection of parameters. There is a general freedom of choice associated with the number of parameters one would like to assign to the elements $\mathcal{C}$. We index the rows with fly number $f$, channel location $c$, and the conscious arousal state $\psi$. That is, the $(i,j,k)$th element is
\begin{gather}
     [\mathcal{C}]_{(i,j,k)} = C^{(i,j,k)}_\mu.
\end{gather}
In other words, it is the $i$th fly's $j$th channel in $k$th condition. Thus, $\mathcal{C}$ has length of $|f|\times|c|\times|\psi|=390$. Each $C_{\mu}$ in this vector is a function of $\lambda$. 

\begin{table*}[t]
\footnotesize
\centering
\begin{tabular}{c|ll|ll|ll}
\hline \hline
$Q$ & \multicolumn{2}{c|}{$1^{st}$ Order} & \multicolumn{2}{c|}{$2^{nd}$ Order} & \multicolumn{2}{c}{$3^{rd}$ Order} \\ \hline
$C_{\mu}^+$ & $\lambda:\;\chi^2(1)=443.64$ & $\;\;p<10^{-16}$ & $\lambda c:\;\chi^2(14)=42.876$ & $\;\;p = 8.97\times 10^{-5}$ & $\lambda c \psi:\; \chi^2(14)=24.00$ & $\;\;p=0.0458$ \\
{} & {} & {} & $\lambda \psi:\;\chi^2(1)=14.63$ & $\;\;p=1.31\times 10^{-4}$ & {} & {} \\
&&&&&& \\
$\Delta C_{\mu}^{+}$ & $\lambda:\; \chi^2(1)=20.97$ & $\;\;p=4.65\times 10^{-6}$ & $\lambda c:\; \chi^2(14)=37.19$ & $\;\;p=6.90\times 10^{-4}$ & {} & {} \\ 
&&&&&& \\
$\Xi$ & $\psi:\; \chi^2(1)=4.870$ & $\;\;p=0.0273$ & $\lambda \psi:\; \chi^2(1)=5.565$ & $\;\;p=0.0183$ & $\lambda c \psi:\; \chi^2(14)=31.79$ & $\;\;p=4.29\times 10^{-3}$ \\
{} & $\lambda:\; \chi^2(1)=6.725$ & $\;\;p=9.51\times 10^{-3}$ & {} & {} & {} & {}  \\
&&&&&& \\
$d$ & $\psi:\; \chi^2(1) = 5.896$ & $\;\;p=0.0152$ & $\lambda \psi:\; \chi^2(1)=6.119$ & $\;\;p=0.0134$ & {} & {} \\
{} & $\lambda:\;\chi^2(1)=460.5$ & $\;\;p<10^{-16}$ & $\lambda c:\;\chi^2(14)=35.86$ & $\;\;p=1.09\times 10^{-3}$ & {} & {} \\
&&&&&& \\
$\mathcal{D}_{KLS}$ & $\lambda:\; \chi^2(1)=127.4$ & $\;\;p<10^{-16}$ & $\lambda c:\; \chi^2(14)=85.81$ & $p<10^{-16}$ & {} & {} \\
{} & {} & {} & $\lambda \psi:\; \chi^2(1)=127.4$ & $\;\;p<10^{-16}$ & {} & {} \\
\hline \hline
\end{tabular}
\caption{Significant $\chi^2$ and $p$ values of effects of channel $c$, memory length $\lambda$, and condition $\psi$, for informational quantities $Q$ obtained via LME analysis. First order effects correspond to significant channel, memory, or condition responses on an informational quantity, while second and third-order effects correspond to interactions between these effects. $\chi^2$ values are reported with $n-1$ degrees of freedom in the parentheses, corresponding to the number of effects removed under the null model, described in Sec.~\ref{sec:Methods}}
\label{tab:LMEresults}
\end{table*}

The matrix $\v{F}$ introducing the set of fixed effects $\mathcal{F} = \{\lambda, c, \psi, \lambda c, \lambda \psi, c \psi, \lambda c \psi \}$ into the model (known in the context of general linear models as the \emph{design matrix}) can then be represented as $\v{F} = (\v{F}_1,\dots, \v{F}_{13})^T$, with each element corresponding to the design matrix of a specific fly. These individual fly response matrices can be explicitly expressed as
\begin{gather}
\v{F}_f =
\begin{pmatrix}
     \vec{\gv{\lambda}} & \v{D} & \vec{\gv\Psi}_W &
     \lambda\v{D} & \lambda\vec{\gv{\Psi}}_W & \v{D}_{\Psi_W} & \lambda\v{D}_{\Psi_W}\\
     \vec{\gv{\lambda}} & \v{D} & \vec{\gv\Psi}_A &
     \lambda\v{D} & \lambda\vec{\gv{\Psi}}_A & \v{D}_{\Psi_A} & \lambda\v{D}_{\Psi_A}
\end{pmatrix},
\end{gather}
where $\vec{\gv{\lambda}} = (\lambda,\ldots,\lambda)^{T}$ and $\vec{\gv{\Psi}}_X = (\Psi_X,\ldots,\Psi_X)^{T}$ are column vectors of length $15$ containing the predictor variables of memory length and level of conscious arousal respectively,  $\v{D}$ is the $15\times 15$ identity matrix which ``selects out" the channel of interest, $\v{D}_{\Psi_X} = \text{diag}(\Psi_X,\ldots, \Psi_X)$ is the $15\times 15$ matrix which ``selects out" the condition of interest correlated with the level of conscious arousal, where
\begin{gather}
  \Psi_{W (A)} =
  \begin{cases}
    1 & \text{if $\psi=$ wakeful (anaesthetised)} \\
    0 & \text{otherwise}.
  \end{cases}
\end{gather}

In a similar fashion, the expression for the matrix containing the random effects $\v{R}$ can be determined. For the case of our study, we only consider random effects arising due to correlations between channels within a specific fly. The result of this is an adjustment to the intercept of the linear model for each fly and channel combination. Therefore, the random effects matrix $\v{R}$ is simply an identity matrix of dimension $390$.
The accompanying elements of the random effects vector $\v{b}$ consist of regression coefficients $b_{fc}$ describing the strength of each intercept adjustment. 

The execution of the LME analysis which included coefficient fitting, and log-likelihood estimations was facilitated by running \texttt{fitlme.m} in MATLAB R2108b.
\bibliography{references}

\begin{thebibliography}{74}%
\makeatletter
\providecommand \@ifxundefined [1]{%
 \@ifx{#1\undefined}
}%
\providecommand \@ifnum [1]{%
 \ifnum #1\expandafter \@firstoftwo
 \else \expandafter \@secondoftwo
 \fi
}%
\providecommand \@ifx [1]{%
 \ifx #1\expandafter \@firstoftwo
 \else \expandafter \@secondoftwo
 \fi
}%
\providecommand \natexlab [1]{#1}%
\providecommand \enquote  [1]{``#1''}%
\providecommand \bibnamefont  [1]{#1}%
\providecommand \bibfnamefont [1]{#1}%
\providecommand \citenamefont [1]{#1}%
\providecommand \href@noop [0]{\@secondoftwo}%
\providecommand \href [0]{\begingroup \@sanitize@url \@href}%
\providecommand \@href[1]{\@@startlink{#1}\@@href}%
\providecommand \@@href[1]{\endgroup#1\@@endlink}%
\providecommand \@sanitize@url [0]{\catcode `\\12\catcode `\$12\catcode
  `\&12\catcode `\#12\catcode `\^12\catcode `\_12\catcode `\%12\relax}%
\providecommand \@@startlink[1]{}%
\providecommand \@@endlink[0]{}%
\providecommand \url  [0]{\begingroup\@sanitize@url \@url }%
\providecommand \@url [1]{\endgroup\@href {#1}{\urlprefix }}%
\providecommand \urlprefix  [0]{URL }%
\providecommand \Eprint [0]{\href }%
\providecommand \doibase [0]{http://dx.doi.org/}%
\providecommand \selectlanguage [0]{\@gobble}%
\providecommand \bibinfo  [0]{\@secondoftwo}%
\providecommand \bibfield  [0]{\@secondoftwo}%
\providecommand \translation [1]{[#1]}%
\providecommand \BibitemOpen [0]{}%
\providecommand \bibitemStop [0]{}%
\providecommand \bibitemNoStop [0]{.\EOS\space}%
\providecommand \EOS [0]{\spacefactor3000\relax}%
\providecommand \BibitemShut  [1]{\csname bibitem#1\endcsname}%
\let\auto@bib@innerbib\@empty
\bibitem [{\citenamefont {Thurner}\ \emph {et~al.}(2018)\citenamefont
  {Thurner}, \citenamefont {Klimek},\ and\ \citenamefont
  {Hanel}}]{Thurner2018}%
  \BibitemOpen
  \bibfield  {author} {\bibinfo {author} {\bibfnamefont {S.}~\bibnamefont
  {Thurner}}, \bibinfo {author} {\bibfnamefont {P.}~\bibnamefont {Klimek}}, \
  and\ \bibinfo {author} {\bibfnamefont {R.}~\bibnamefont {Hanel}},\ }\href
  {\doibase 10.1093/oso/9780198821939.001.0001} {\emph {\bibinfo {title}
  {Introduction to the Theory of Complex Systems}}}\ (\bibinfo  {publisher}
  {Oxford University Press},\ \bibinfo {address} {Oxford},\ \bibinfo {year}
  {2018})\BibitemShut {NoStop}%
\bibitem [{\citenamefont {Gazzaniga}(2009)}]{CognitiveNeurosciences}%
  \BibitemOpen
  \bibfield  {author} {\bibinfo {author} {\bibfnamefont {M.~S.}\ \bibnamefont
  {Gazzaniga}},\ }\href@noop {} {\emph {\bibinfo {title} {The Cognitive
  Neurosciences}}},\ \bibinfo {edition} {4th}\ ed.\ (\bibinfo  {publisher} {The
  MIT Press},\ \bibinfo {year} {2009})\BibitemShut {NoStop}%
\bibitem [{\citenamefont {Boly}\ \emph {et~al.}(2013)\citenamefont {Boly},
  \citenamefont {Seth}, \citenamefont {Wilke}, \citenamefont {Ingmundson},
  \citenamefont {Baars}, \citenamefont {Laureys}, \citenamefont {Edelman},\
  and\ \citenamefont {Tsuchiya}}]{Boly2013}%
  \BibitemOpen
  \bibfield  {author} {\bibinfo {author} {\bibfnamefont {M.}~\bibnamefont
  {Boly}}, \bibinfo {author} {\bibfnamefont {A.}~\bibnamefont {Seth}}, \bibinfo
  {author} {\bibfnamefont {M.}~\bibnamefont {Wilke}}, \bibinfo {author}
  {\bibfnamefont {P.}~\bibnamefont {Ingmundson}}, \bibinfo {author}
  {\bibfnamefont {B.}~\bibnamefont {Baars}}, \bibinfo {author} {\bibfnamefont
  {S.}~\bibnamefont {Laureys}}, \bibinfo {author} {\bibfnamefont
  {D.}~\bibnamefont {Edelman}}, \ and\ \bibinfo {author} {\bibfnamefont
  {N.}~\bibnamefont {Tsuchiya}},\ }\href {\doibase 10.3389/fpsyg.2013.00625}
  {\bibfield  {journal} {\bibinfo  {journal} {Front. Psychol.}\ }\textbf
  {\bibinfo {volume} {4}},\ \bibinfo {pages} {625} (\bibinfo {year}
  {2013})}\BibitemShut {NoStop}%
\bibitem [{\citenamefont {Laureys}\ and\ \citenamefont
  {Schiff}(2012)}]{Laureys2012}%
  \BibitemOpen
  \bibfield  {author} {\bibinfo {author} {\bibfnamefont {S.}~\bibnamefont
  {Laureys}}\ and\ \bibinfo {author} {\bibfnamefont {N.~D.}\ \bibnamefont
  {Schiff}},\ }\href {\doibase 10.1016/j.neuroimage.2011.12.041} {\bibfield
  {journal} {\bibinfo  {journal} {Neuroimage}\ }\textbf {\bibinfo {volume}
  {61}},\ \bibinfo {pages} {478} (\bibinfo {year} {2012})}\BibitemShut
  {NoStop}%
\bibitem [{\citenamefont {Laureys}\ \emph {et~al.}(2015)\citenamefont
  {Laureys}, \citenamefont {Gosseries},\ and\ \citenamefont
  {Tononi}}]{NeurologyOfConsciousness}%
  \BibitemOpen
  \bibfield  {author} {\bibinfo {author} {\bibfnamefont {S.}~\bibnamefont
  {Laureys}}, \bibinfo {author} {\bibfnamefont {O.}~\bibnamefont {Gosseries}},
  \ and\ \bibinfo {author} {\bibfnamefont {G.}~\bibnamefont {Tononi}},\
  }\href@noop {} {\emph {\bibinfo {title} {The Neurology of Consciousness}}},\
  \bibinfo {edition} {2nd}\ ed.\ (\bibinfo  {publisher} {Elsevier},\ \bibinfo
  {year} {2015})\BibitemShut {NoStop}%
\bibitem [{\citenamefont {Engemann}\ \emph {et~al.}(2018)\citenamefont
  {Engemann}, \citenamefont {Raimondo}, \citenamefont {King}, \citenamefont
  {Rohaut}, \citenamefont {Louppe}, \citenamefont {Faugeras}, \citenamefont
  {Annen}, \citenamefont {Cassol}, \citenamefont {Gosseries}, \citenamefont
  {Fernandez-Slezak} \emph {et~al.}}]{Engemann2018}%
  \BibitemOpen
  \bibfield  {author} {\bibinfo {author} {\bibfnamefont {D.~A.}\ \bibnamefont
  {Engemann}}, \bibinfo {author} {\bibfnamefont {F.}~\bibnamefont {Raimondo}},
  \bibinfo {author} {\bibfnamefont {J.-R.}\ \bibnamefont {King}}, \bibinfo
  {author} {\bibfnamefont {B.}~\bibnamefont {Rohaut}}, \bibinfo {author}
  {\bibfnamefont {G.}~\bibnamefont {Louppe}}, \bibinfo {author} {\bibfnamefont
  {F.}~\bibnamefont {Faugeras}}, \bibinfo {author} {\bibfnamefont
  {J.}~\bibnamefont {Annen}}, \bibinfo {author} {\bibfnamefont
  {H.}~\bibnamefont {Cassol}}, \bibinfo {author} {\bibfnamefont
  {O.}~\bibnamefont {Gosseries}}, \bibinfo {author} {\bibfnamefont
  {D.}~\bibnamefont {Fernandez-Slezak}},  \emph {et~al.},\ }\href {\doibase
  10.1093/brain/awy251} {\bibfield  {journal} {\bibinfo  {journal} {Brain}\
  }\textbf {\bibinfo {volume} {141}},\ \bibinfo {pages} {3179} (\bibinfo {year}
  {2018})}\BibitemShut {NoStop}%
\bibitem [{\citenamefont {Sitt}\ \emph {et~al.}(2014)\citenamefont {Sitt},
  \citenamefont {King}, \citenamefont {El~Karoui}, \citenamefont {Rohaut},
  \citenamefont {Faugeras}, \citenamefont {Gramfort}, \citenamefont {Cohen},
  \citenamefont {Sigman}, \citenamefont {Dehaene},\ and\ \citenamefont
  {Naccache}}]{Sitt2014}%
  \BibitemOpen
  \bibfield  {author} {\bibinfo {author} {\bibfnamefont {J.~D.}\ \bibnamefont
  {Sitt}}, \bibinfo {author} {\bibfnamefont {J.-R.}\ \bibnamefont {King}},
  \bibinfo {author} {\bibfnamefont {I.}~\bibnamefont {El~Karoui}}, \bibinfo
  {author} {\bibfnamefont {B.}~\bibnamefont {Rohaut}}, \bibinfo {author}
  {\bibfnamefont {F.}~\bibnamefont {Faugeras}}, \bibinfo {author}
  {\bibfnamefont {A.}~\bibnamefont {Gramfort}}, \bibinfo {author}
  {\bibfnamefont {L.}~\bibnamefont {Cohen}}, \bibinfo {author} {\bibfnamefont
  {M.}~\bibnamefont {Sigman}}, \bibinfo {author} {\bibfnamefont
  {S.}~\bibnamefont {Dehaene}}, \ and\ \bibinfo {author} {\bibfnamefont
  {L.}~\bibnamefont {Naccache}},\ }\href {\doibase 10.1093/brain/awu141}
  {\bibfield  {journal} {\bibinfo  {journal} {Brain}\ }\textbf {\bibinfo
  {volume} {137}},\ \bibinfo {pages} {2258} (\bibinfo {year}
  {2014})}\BibitemShut {NoStop}%
\bibitem [{\citenamefont {Massimini}\ \emph {et~al.}(2005)\citenamefont
  {Massimini}, \citenamefont {Ferrarelli}, \citenamefont {Huber}, \citenamefont
  {Esser}, \citenamefont {Singh},\ and\ \citenamefont
  {Tononi}}]{Massimini2005}%
  \BibitemOpen
  \bibfield  {author} {\bibinfo {author} {\bibfnamefont {M.}~\bibnamefont
  {Massimini}}, \bibinfo {author} {\bibfnamefont {F.}~\bibnamefont
  {Ferrarelli}}, \bibinfo {author} {\bibfnamefont {R.}~\bibnamefont {Huber}},
  \bibinfo {author} {\bibfnamefont {S.~K.}\ \bibnamefont {Esser}}, \bibinfo
  {author} {\bibfnamefont {H.}~\bibnamefont {Singh}}, \ and\ \bibinfo {author}
  {\bibfnamefont {G.}~\bibnamefont {Tononi}},\ }\href {\doibase
  10.1126/science.1117256} {\bibfield  {journal} {\bibinfo  {journal}
  {Science}\ }\textbf {\bibinfo {volume} {309}},\ \bibinfo {pages} {2228}
  (\bibinfo {year} {2005})}\BibitemShut {NoStop}%
\bibitem [{\citenamefont {Casali}\ \emph {et~al.}(2013)\citenamefont {Casali},
  \citenamefont {Gosseries}, \citenamefont {Rosanova}, \citenamefont {Boly},
  \citenamefont {Sarasso}, \citenamefont {Casali}, \citenamefont {Casarotto},
  \citenamefont {Bruno}, \citenamefont {Laureys}, \citenamefont {Tononi},\ and\
  \citenamefont {Massimini}}]{Casali2013}%
  \BibitemOpen
  \bibfield  {author} {\bibinfo {author} {\bibfnamefont {A.~G.}\ \bibnamefont
  {Casali}}, \bibinfo {author} {\bibfnamefont {O.}~\bibnamefont {Gosseries}},
  \bibinfo {author} {\bibfnamefont {M.}~\bibnamefont {Rosanova}}, \bibinfo
  {author} {\bibfnamefont {M.}~\bibnamefont {Boly}}, \bibinfo {author}
  {\bibfnamefont {S.}~\bibnamefont {Sarasso}}, \bibinfo {author} {\bibfnamefont
  {K.~R.}\ \bibnamefont {Casali}}, \bibinfo {author} {\bibfnamefont
  {S.}~\bibnamefont {Casarotto}}, \bibinfo {author} {\bibfnamefont {M.-A.}\
  \bibnamefont {Bruno}}, \bibinfo {author} {\bibfnamefont {S.}~\bibnamefont
  {Laureys}}, \bibinfo {author} {\bibfnamefont {G.}~\bibnamefont {Tononi}}, \
  and\ \bibinfo {author} {\bibfnamefont {M.}~\bibnamefont {Massimini}},\ }\href
  {\doibase 10.1126/scitranslmed.3006294} {\bibfield  {journal} {\bibinfo
  {journal} {Sci. Transl. Med.}\ }\textbf {\bibinfo {volume} {5}},\ \bibinfo
  {pages} {198ra105} (\bibinfo {year} {2013})}\BibitemShut {NoStop}%
\bibitem [{\citenamefont {Casarotto}\ \emph {et~al.}(2016)\citenamefont
  {Casarotto}, \citenamefont {Comanducci}, \citenamefont {Rosanova},
  \citenamefont {Sarasso}, \citenamefont {Fecchio}, \citenamefont {Napolitani},
  \citenamefont {Pigorini}, \citenamefont {Casali}, \citenamefont {Trimarchi},
  \citenamefont {Boly}, \citenamefont {Gosseries}, \citenamefont {Bodart},
  \citenamefont {Curto}, \citenamefont {Landi}, \citenamefont {Mariotti},
  \citenamefont {Devalle}, \citenamefont {Laureys}, \citenamefont {Tononi},\
  and\ \citenamefont {Massimini}}]{Casarotto2016}%
  \BibitemOpen
  \bibfield  {author} {\bibinfo {author} {\bibfnamefont {S.}~\bibnamefont
  {Casarotto}}, \bibinfo {author} {\bibfnamefont {A.}~\bibnamefont
  {Comanducci}}, \bibinfo {author} {\bibfnamefont {M.}~\bibnamefont
  {Rosanova}}, \bibinfo {author} {\bibfnamefont {S.}~\bibnamefont {Sarasso}},
  \bibinfo {author} {\bibfnamefont {M.}~\bibnamefont {Fecchio}}, \bibinfo
  {author} {\bibfnamefont {M.}~\bibnamefont {Napolitani}}, \bibinfo {author}
  {\bibfnamefont {A.}~\bibnamefont {Pigorini}}, \bibinfo {author}
  {\bibfnamefont {A.~G.}\ \bibnamefont {Casali}}, \bibinfo {author}
  {\bibfnamefont {P.~D.}\ \bibnamefont {Trimarchi}}, \bibinfo {author}
  {\bibfnamefont {M.}~\bibnamefont {Boly}}, \bibinfo {author} {\bibfnamefont
  {O.}~\bibnamefont {Gosseries}}, \bibinfo {author} {\bibfnamefont
  {O.}~\bibnamefont {Bodart}}, \bibinfo {author} {\bibfnamefont
  {F.}~\bibnamefont {Curto}}, \bibinfo {author} {\bibfnamefont
  {C.}~\bibnamefont {Landi}}, \bibinfo {author} {\bibfnamefont
  {M.}~\bibnamefont {Mariotti}}, \bibinfo {author} {\bibfnamefont
  {G.}~\bibnamefont {Devalle}}, \bibinfo {author} {\bibfnamefont
  {S.}~\bibnamefont {Laureys}}, \bibinfo {author} {\bibfnamefont
  {G.}~\bibnamefont {Tononi}}, \ and\ \bibinfo {author} {\bibfnamefont
  {M.}~\bibnamefont {Massimini}},\ }\href {\doibase 10.1002/ana.24779}
  {\bibfield  {journal} {\bibinfo  {journal} {Ann. Neurol.}\ }\textbf {\bibinfo
  {volume} {80}},\ \bibinfo {pages} {718} (\bibinfo {year} {2016})}\BibitemShut
  {NoStop}%
\bibitem [{\citenamefont {Tononi}(2004)}]{Tononi2004}%
  \BibitemOpen
  \bibfield  {author} {\bibinfo {author} {\bibfnamefont {G.}~\bibnamefont
  {Tononi}},\ }\href {\doibase 10.1186/1471-2202-5-42} {\bibfield  {journal}
  {\bibinfo  {journal} {BMC Neurosci.}\ }\textbf {\bibinfo {volume} {5}},\
  \bibinfo {pages} {42} (\bibinfo {year} {2004})}\BibitemShut {NoStop}%
\bibitem [{\citenamefont {Tononi}\ \emph {et~al.}(2016)\citenamefont {Tononi},
  \citenamefont {Boly}, \citenamefont {Massimini},\ and\ \citenamefont
  {Koch}}]{Tononi2016}%
  \BibitemOpen
  \bibfield  {author} {\bibinfo {author} {\bibfnamefont {G.}~\bibnamefont
  {Tononi}}, \bibinfo {author} {\bibfnamefont {M.}~\bibnamefont {Boly}},
  \bibinfo {author} {\bibfnamefont {M.}~\bibnamefont {Massimini}}, \ and\
  \bibinfo {author} {\bibfnamefont {C.}~\bibnamefont {Koch}},\ }\href {\doibase
  10.1038/nrn.2016.44} {\bibfield  {journal} {\bibinfo  {journal} {Nat. Rev.
  Neurosci.}\ }\textbf {\bibinfo {volume} {17}},\ \bibinfo {pages} {450}
  (\bibinfo {year} {2016})}\BibitemShut {NoStop}%
\bibitem [{\citenamefont {Oizumi}\ \emph {et~al.}(2014)\citenamefont {Oizumi},
  \citenamefont {Albantakis},\ and\ \citenamefont {Tononi}}]{Oizumi2014}%
  \BibitemOpen
  \bibfield  {author} {\bibinfo {author} {\bibfnamefont {M.}~\bibnamefont
  {Oizumi}}, \bibinfo {author} {\bibfnamefont {L.}~\bibnamefont {Albantakis}},
  \ and\ \bibinfo {author} {\bibfnamefont {G.}~\bibnamefont {Tononi}},\ }\href
  {\doibase 10.1371/journal.pcbi.1003588} {\bibfield  {journal} {\bibinfo
  {journal} {PLOS Comput. Biol.}\ }\textbf {\bibinfo {volume} {10}},\ \bibinfo
  {pages} {e1003588} (\bibinfo {year} {2014})}\BibitemShut {NoStop}%
\bibitem [{\citenamefont {Mediano}\ \emph {et~al.}(2019)\citenamefont
  {Mediano}, \citenamefont {Seth},\ and\ \citenamefont
  {Barrett}}]{Mediano2019}%
  \BibitemOpen
  \bibfield  {author} {\bibinfo {author} {\bibfnamefont {P.~A.~M.}\
  \bibnamefont {Mediano}}, \bibinfo {author} {\bibfnamefont {A.~K.}\
  \bibnamefont {Seth}}, \ and\ \bibinfo {author} {\bibfnamefont {A.~B.}\
  \bibnamefont {Barrett}},\ }\href {\doibase 10.3390/e21010017} {\bibfield
  {journal} {\bibinfo  {journal} {Entropy}\ }\textbf {\bibinfo {volume} {21}},\
  \bibinfo {pages} {17} (\bibinfo {year} {2019})}\BibitemShut {NoStop}%
\bibitem [{\citenamefont {Barrett}\ and\ \citenamefont
  {Seth}(2011)}]{Barrett2011}%
  \BibitemOpen
  \bibfield  {author} {\bibinfo {author} {\bibfnamefont {A.~B.}\ \bibnamefont
  {Barrett}}\ and\ \bibinfo {author} {\bibfnamefont {A.~K.}\ \bibnamefont
  {Seth}},\ }\href {\doibase 10.1371/journal.pcbi.1001052} {\bibfield
  {journal} {\bibinfo  {journal} {PLOS Comput. Biol.}\ }\textbf {\bibinfo
  {volume} {7}},\ \bibinfo {pages} {e1001052} (\bibinfo {year}
  {2011})}\BibitemShut {NoStop}%
\bibitem [{\citenamefont {Tegmark}(2016)}]{Tegmark2016}%
  \BibitemOpen
  \bibfield  {author} {\bibinfo {author} {\bibfnamefont {M.}~\bibnamefont
  {Tegmark}},\ }\href {\doibase 10.1371/journal.pcbi.1005123} {\bibfield
  {journal} {\bibinfo  {journal} {PLOS Comput. Biol.}\ }\textbf {\bibinfo
  {volume} {12}},\ \bibinfo {pages} {e1005123} (\bibinfo {year}
  {2016})}\BibitemShut {NoStop}%
\bibitem [{\citenamefont {Oizumi}\ \emph
  {et~al.}(2016{\natexlab{a}})\citenamefont {Oizumi}, \citenamefont
  {Tsuchiya},\ and\ \citenamefont {Amari}}]{Oizumi2016PNAS}%
  \BibitemOpen
  \bibfield  {author} {\bibinfo {author} {\bibfnamefont {M.}~\bibnamefont
  {Oizumi}}, \bibinfo {author} {\bibfnamefont {N.}~\bibnamefont {Tsuchiya}}, \
  and\ \bibinfo {author} {\bibfnamefont {S.}~\bibnamefont {Amari}},\ }\href
  {\doibase 10.1073/pnas.1603583113} {\bibfield  {journal} {\bibinfo  {journal}
  {Proc. Natl. Acad. Sci.}\ }\textbf {\bibinfo {volume} {113}},\ \bibinfo
  {pages} {14817} (\bibinfo {year} {2016}{\natexlab{a}})}\BibitemShut {NoStop}%
\bibitem [{\citenamefont {Oizumi}\ \emph
  {et~al.}(2016{\natexlab{b}})\citenamefont {Oizumi}, \citenamefont {Amari},
  \citenamefont {Yanagawa}, \citenamefont {Fujii},\ and\ \citenamefont
  {Tsuchiya}}]{Oizumi2016PLOS}%
  \BibitemOpen
  \bibfield  {author} {\bibinfo {author} {\bibfnamefont {M.}~\bibnamefont
  {Oizumi}}, \bibinfo {author} {\bibfnamefont {S.}~\bibnamefont {Amari}},
  \bibinfo {author} {\bibfnamefont {T.}~\bibnamefont {Yanagawa}}, \bibinfo
  {author} {\bibfnamefont {N.}~\bibnamefont {Fujii}}, \ and\ \bibinfo {author}
  {\bibfnamefont {N.}~\bibnamefont {Tsuchiya}},\ }\href {\doibase
  10.1371/journal.pcbi.1004654} {\bibfield  {journal} {\bibinfo  {journal}
  {PLOS Comput. Biol.}\ }\textbf {\bibinfo {volume} {12}},\ \bibinfo {pages}
  {1} (\bibinfo {year} {2016}{\natexlab{b}})}\BibitemShut {NoStop}%
\bibitem [{\citenamefont {Buzsaki}(2006)}]{Buzsaki2006}%
  \BibitemOpen
  \bibfield  {author} {\bibinfo {author} {\bibfnamefont {G.}~\bibnamefont
  {Buzsaki}},\ }\href {\doibase 10.1093/acprof:oso/9780195301069.001.0001}
  {\emph {\bibinfo {title} {Rhythms of the Brain}}}\ (\bibinfo  {publisher}
  {Oxford University Press},\ \bibinfo {year} {2006})\BibitemShut {NoStop}%
\bibitem [{\citenamefont {Crutchfield}\ and\ \citenamefont
  {Young}(1989)}]{CrutchPRL1989}%
  \BibitemOpen
  \bibfield  {author} {\bibinfo {author} {\bibfnamefont {J.~P.}\ \bibnamefont
  {Crutchfield}}\ and\ \bibinfo {author} {\bibfnamefont {K.}~\bibnamefont
  {Young}},\ }\href {\doibase 10.1103/PhysRevLett.63.105} {\bibfield  {journal}
  {\bibinfo  {journal} {Phys. Rev. Lett.}\ }\textbf {\bibinfo {volume} {63}},\
  \bibinfo {pages} {105} (\bibinfo {year} {1989})}\BibitemShut {NoStop}%
\bibitem [{\citenamefont {Travers}\ and\ \citenamefont
  {Crutchfield}(2011)}]{epsilonMachines2}%
  \BibitemOpen
  \bibfield  {author} {\bibinfo {author} {\bibfnamefont {N.~F.}\ \bibnamefont
  {Travers}}\ and\ \bibinfo {author} {\bibfnamefont {J.~P.}\ \bibnamefont
  {Crutchfield}},\ }\href {https://arxiv.org/abs/1111.4500} {\bibfield
  {journal} {\bibinfo  {journal} {arXiv:1111.4500}\ } (\bibinfo {year}
  {2011})}\BibitemShut {NoStop}%
\bibitem [{\citenamefont {Crutchfield}(2017)}]{CrutcharXiv2017}%
  \BibitemOpen
  \bibfield  {author} {\bibinfo {author} {\bibfnamefont {J.~P.}\ \bibnamefont
  {Crutchfield}},\ }\href {https://arxiv.org/abs/1710.06832} {\bibfield
  {journal} {\bibinfo  {journal} {arXiv:1710.06832}\ } (\bibinfo {year}
  {2017})}\BibitemShut {NoStop}%
\bibitem [{\citenamefont {Heinz~Haslinger}\ \emph {et~al.}(2009)\citenamefont
  {Heinz~Haslinger}, \citenamefont {Lisa~Klinkner},\ and\ \citenamefont
  {Rohilla~Shalizi}}]{Haslinger2009}%
  \BibitemOpen
  \bibfield  {author} {\bibinfo {author} {\bibfnamefont {R.}~\bibnamefont
  {Heinz~Haslinger}}, \bibinfo {author} {\bibfnamefont {K.}~\bibnamefont
  {Lisa~Klinkner}}, \ and\ \bibinfo {author} {\bibfnamefont {C.}~\bibnamefont
  {Rohilla~Shalizi}},\ }\href {\doibase 10.1162/neco.2009.12-07-678} {\bibfield
   {journal} {\bibinfo  {journal} {Neural Comput.}\ }\textbf {\bibinfo {volume}
  {22}},\ \bibinfo {pages} {121} (\bibinfo {year} {2009})}\BibitemShut
  {NoStop}%
\bibitem [{\citenamefont {Klinkner}\ \emph {et~al.}(2006)\citenamefont
  {Klinkner}, \citenamefont {Shalizi},\ and\ \citenamefont
  {Camperi}}]{Klinkner2006}%
  \BibitemOpen
  \bibfield  {author} {\bibinfo {author} {\bibfnamefont {K.}~\bibnamefont
  {Klinkner}}, \bibinfo {author} {\bibfnamefont {C.}~\bibnamefont {Shalizi}}, \
  and\ \bibinfo {author} {\bibfnamefont {M.}~\bibnamefont {Camperi}},\ }in\
  \href@noop {} {\emph {\bibinfo {booktitle} {Advances in neural information
  processing systems}}}\ (\bibinfo {year} {2006})\ pp.\ \bibinfo {pages}
  {667--674}\BibitemShut {NoStop}%
\bibitem [{\citenamefont {Shalizi}\ and\ \citenamefont
  {Klinkner}(2004)}]{CSSR2}%
  \BibitemOpen
  \bibfield  {author} {\bibinfo {author} {\bibfnamefont {C.~R.}\ \bibnamefont
  {Shalizi}}\ and\ \bibinfo {author} {\bibfnamefont {K.~L.}\ \bibnamefont
  {Klinkner}},\ }in\ \href {http://arxiv.org/abs/cs.LG/0406011} {\emph
  {\bibinfo {booktitle} {Uncertainty in Artificial Intelligence: Proceedings of
  the Twentieth Conference (UAI 2004)}}},\ \bibinfo {editor} {edited by\
  \bibinfo {editor} {\bibfnamefont {M.}~\bibnamefont {Chickering}}\ and\
  \bibinfo {editor} {\bibfnamefont {J.~Y.}\ \bibnamefont {Halpern}}}\ (\bibinfo
   {publisher} {AUAI Press},\ \bibinfo {address} {Arlington, Virginia},\
  \bibinfo {year} {2004})\ pp.\ \bibinfo {pages} {504--511}\BibitemShut
  {NoStop}%
\bibitem [{\citenamefont {Varn}\ and\ \citenamefont
  {Crutchfield}(2004)}]{Varn2004}%
  \BibitemOpen
  \bibfield  {author} {\bibinfo {author} {\bibfnamefont {D.}~\bibnamefont
  {Varn}}\ and\ \bibinfo {author} {\bibfnamefont {J.}~\bibnamefont
  {Crutchfield}},\ }\href {\doibase 10.1016/j.physleta.2004.02.077} {\bibfield
  {journal} {\bibinfo  {journal} {Phys. Lett. A}\ }\textbf {\bibinfo {volume}
  {324}},\ \bibinfo {pages} {299 } (\bibinfo {year} {2004})}\BibitemShut
  {NoStop}%
\bibitem [{\citenamefont {Boschetti}(2008)}]{Boschetti2008}%
  \BibitemOpen
  \bibfield  {author} {\bibinfo {author} {\bibfnamefont {F.}~\bibnamefont
  {Boschetti}},\ }\href {\doibase https://doi.org/10.1016/j.ecocom.2007.09.002}
  {\bibfield  {journal} {\bibinfo  {journal} {Ecol. Complex.}\ }\textbf
  {\bibinfo {volume} {5}},\ \bibinfo {pages} {37 } (\bibinfo {year}
  {2008})}\BibitemShut {NoStop}%
\bibitem [{\citenamefont {Park}\ \emph {et~al.}(2007)\citenamefont {Park},
  \citenamefont {Lee}, \citenamefont {Yang}, \citenamefont {Jo},\ and\
  \citenamefont {Moon}}]{Park2007}%
  \BibitemOpen
  \bibfield  {author} {\bibinfo {author} {\bibfnamefont {J.~B.}\ \bibnamefont
  {Park}}, \bibinfo {author} {\bibfnamefont {J.~W.}\ \bibnamefont {Lee}},
  \bibinfo {author} {\bibfnamefont {J.-S.}\ \bibnamefont {Yang}}, \bibinfo
  {author} {\bibfnamefont {H.-H.}\ \bibnamefont {Jo}}, \ and\ \bibinfo {author}
  {\bibfnamefont {H.-T.}\ \bibnamefont {Moon}},\ }\href {\doibase
  10.1016/j.physa.2006.12.042} {\bibfield  {journal} {\bibinfo  {journal}
  {Physica A}\ }\textbf {\bibinfo {volume} {379}},\ \bibinfo {pages} {179}
  (\bibinfo {year} {2007})}\BibitemShut {NoStop}%
\bibitem [{\citenamefont {Cohen}\ \emph {et~al.}(2016)\citenamefont {Cohen},
  \citenamefont {Zalucki}, \citenamefont {van Swinderen},\ and\ \citenamefont
  {Tsuchiya}}]{CohenEneuro2016}%
  \BibitemOpen
  \bibfield  {author} {\bibinfo {author} {\bibfnamefont {D.}~\bibnamefont
  {Cohen}}, \bibinfo {author} {\bibfnamefont {O.~H.}\ \bibnamefont {Zalucki}},
  \bibinfo {author} {\bibfnamefont {B.}~\bibnamefont {van Swinderen}}, \ and\
  \bibinfo {author} {\bibfnamefont {N.}~\bibnamefont {Tsuchiya}},\ }\href
  {https://dx.doi.org/10.1523/ENEURO.0116-16.2016} {\bibfield  {journal}
  {\bibinfo  {journal} {eNeuro}\ }\textbf {\bibinfo {volume} {3}} (\bibinfo
  {year} {2016})}\BibitemShut {NoStop}%
\bibitem [{\citenamefont {Cohen}\ \emph {et~al.}(2018)\citenamefont {Cohen},
  \citenamefont {van Swinderen},\ and\ \citenamefont {Tsuchiya}}]{CohenEneuro}%
  \BibitemOpen
  \bibfield  {author} {\bibinfo {author} {\bibfnamefont {D.}~\bibnamefont
  {Cohen}}, \bibinfo {author} {\bibfnamefont {B.}~\bibnamefont {van
  Swinderen}}, \ and\ \bibinfo {author} {\bibfnamefont {N.}~\bibnamefont
  {Tsuchiya}},\ }\href {https://dx.doi.org/10.1523/ENEURO.0329-17.2018}
  {\bibfield  {journal} {\bibinfo  {journal} {eNeuro}\ }\textbf {\bibinfo
  {volume} {5}} (\bibinfo {year} {2018})}\BibitemShut {NoStop}%
\bibitem [{\citenamefont {Hohwy}(2013)}]{Hohwy2013}%
  \BibitemOpen
  \bibfield  {author} {\bibinfo {author} {\bibfnamefont {J.}~\bibnamefont
  {Hohwy}},\ }\href {\doibase 10.1093/acprof:oso/9780199682737.001.0001}
  {{\selectlanguage {English}\emph {\bibinfo {title} {The Predictive Mind}}}}\
  (\bibinfo  {publisher} {Oxford University Press},\ \bibinfo {address} {United
  Kingdom},\ \bibinfo {year} {2013})\BibitemShut {NoStop}%
\bibitem [{\citenamefont {Tononi}(2010)}]{Tononi2010}%
  \BibitemOpen
  \bibfield  {author} {\bibinfo {author} {\bibfnamefont {G.}~\bibnamefont
  {Tononi}},\ }\href@noop {} {\bibfield  {journal} {\bibinfo  {journal} {Arch.
  Ital. Biol.}\ }\textbf {\bibinfo {volume} {148}},\ \bibinfo {pages} {299}
  (\bibinfo {year} {2010})}\BibitemShut {NoStop}%
\bibitem [{\citenamefont {Crutchfield}\ \emph {et~al.}(2009)\citenamefont
  {Crutchfield}, \citenamefont {Ellison},\ and\ \citenamefont
  {Mahoney}}]{cryptCrutch}%
  \BibitemOpen
  \bibfield  {author} {\bibinfo {author} {\bibfnamefont {J.~P.}\ \bibnamefont
  {Crutchfield}}, \bibinfo {author} {\bibfnamefont {C.~J.}\ \bibnamefont
  {Ellison}}, \ and\ \bibinfo {author} {\bibfnamefont {J.~R.}\ \bibnamefont
  {Mahoney}},\ }\href {\doibase 10.1103/PhysRevLett.103.094101} {\bibfield
  {journal} {\bibinfo  {journal} {Phys. Rev. Lett.}\ }\textbf {\bibinfo
  {volume} {103}},\ \bibinfo {pages} {094101} (\bibinfo {year}
  {2009})}\BibitemShut {NoStop}%
\bibitem [{\citenamefont {Rabiner}(1989)}]{Rabiner1989}%
  \BibitemOpen
  \bibfield  {author} {\bibinfo {author} {\bibfnamefont {L.~R.}\ \bibnamefont
  {Rabiner}},\ }\href {\doibase 10.1109/5.18626} {\bibfield  {journal}
  {\bibinfo  {journal} {Proc. IEEE}\ }\textbf {\bibinfo {volume} {77}},\
  \bibinfo {pages} {257} (\bibinfo {year} {1989})}\BibitemShut {NoStop}%
\bibitem [{\citenamefont {Doob}(1953)}]{DoobStochastic}%
  \BibitemOpen
  \bibfield  {author} {\bibinfo {author} {\bibfnamefont {J.~L.}\ \bibnamefont
  {Doob}},\ }\href@noop {} {\emph {\bibinfo {title} {Stochastic Processes}}}\
  (\bibinfo  {publisher} {Wiley, New York},\ \bibinfo {year}
  {1953})\BibitemShut {NoStop}%
\bibitem [{\citenamefont {Gu}\ \emph {et~al.}(2012)\citenamefont {Gu},
  \citenamefont {Wiesner}, \citenamefont {Rieper},\ and\ \citenamefont
  {Vedral}}]{Gu2012}%
  \BibitemOpen
  \bibfield  {author} {\bibinfo {author} {\bibfnamefont {M.}~\bibnamefont
  {Gu}}, \bibinfo {author} {\bibfnamefont {K.}~\bibnamefont {Wiesner}},
  \bibinfo {author} {\bibfnamefont {E.}~\bibnamefont {Rieper}}, \ and\ \bibinfo
  {author} {\bibfnamefont {V.}~\bibnamefont {Vedral}},\ }\href {\doibase
  10.1038/ncomms1761} {\bibfield  {journal} {\bibinfo  {journal} {Nat.
  Commun.}\ }\textbf {\bibinfo {volume} {3}},\ \bibinfo {pages} {762} (\bibinfo
  {year} {2012})}\BibitemShut {NoStop}%
\bibitem [{\citenamefont {Gagniuc}(2017)}]{Gagniuc2017}%
  \BibitemOpen
  \bibfield  {author} {\bibinfo {author} {\bibfnamefont {P.~A.}\ \bibnamefont
  {Gagniuc}},\ }\href@noop {} {\emph {\bibinfo {title} {Markov chains: from
  theory to implementation and experimentation}}}\ (\bibinfo  {publisher} {John
  Wiley \& Sons},\ \bibinfo {year} {2017})\BibitemShut {NoStop}%
\bibitem [{\citenamefont {Tino}\ and\ \citenamefont
  {Dorffner}(2001)}]{Tino2001}%
  \BibitemOpen
  \bibfield  {author} {\bibinfo {author} {\bibfnamefont {P.}~\bibnamefont
  {Tino}}\ and\ \bibinfo {author} {\bibfnamefont {G.}~\bibnamefont
  {Dorffner}},\ }\href {\doibase 10.1023/A:1010972803901} {\bibfield  {journal}
  {\bibinfo  {journal} {Mach. Learn.}\ }\textbf {\bibinfo {volume} {45}},\
  \bibinfo {pages} {187} (\bibinfo {year} {2001})}\BibitemShut {NoStop}%
\bibitem [{\citenamefont {Crutchfield}\ and\ \citenamefont
  {Young}(1990)}]{Crutchfield1990}%
  \BibitemOpen
  \bibfield  {author} {\bibinfo {author} {\bibfnamefont {J.~P.}\ \bibnamefont
  {Crutchfield}}\ and\ \bibinfo {author} {\bibfnamefont {K.}~\bibnamefont
  {Young}},\ }\enquote {\bibinfo {title} {Computation at the onset of chaos},}\
  in\ \href@noop {} {\emph {\bibinfo {booktitle} {Entropy, Complexity, and the
  Physics of Information}}},\ Vol.~\bibinfo {volume} {8},\ \bibinfo {editor}
  {edited by\ \bibinfo {editor} {\bibfnamefont {W.}~\bibnamefont {Zurek}}}\
  (\bibinfo  {publisher} {Addison-Wesley, Reading, Massachusetts},\ \bibinfo
  {year} {1990})\ pp.\ \bibinfo {pages} {223--269}\BibitemShut {NoStop}%
\bibitem [{\citenamefont {Massey}(1951)}]{Massey1951}%
  \BibitemOpen
  \bibfield  {author} {\bibinfo {author} {\bibfnamefont {F.~J.}\ \bibnamefont
  {Massey}},\ }\href {\doibase 10.1080/01621459.1951.10500769} {\bibfield
  {journal} {\bibinfo  {journal} {J. Am. Stat. Assoc.}\ }\textbf {\bibinfo
  {volume} {46}},\ \bibinfo {pages} {68} (\bibinfo {year} {1951})}\BibitemShut
  {NoStop}%
\bibitem [{\citenamefont {Hollander}\ \emph {et~al.}(2013)\citenamefont
  {Hollander}, \citenamefont {Wolfe},\ and\ \citenamefont
  {Chicken}}]{Hollander2013}%
  \BibitemOpen
  \bibfield  {author} {\bibinfo {author} {\bibfnamefont {M.}~\bibnamefont
  {Hollander}}, \bibinfo {author} {\bibfnamefont {D.~A.}\ \bibnamefont
  {Wolfe}}, \ and\ \bibinfo {author} {\bibfnamefont {E.}~\bibnamefont
  {Chicken}},\ }\href@noop {} {\emph {\bibinfo {title} {Nonparametric
  statistical methods}}},\ Vol.\ \bibinfo {volume} {751}\ (\bibinfo
  {publisher} {John Wiley \& Sons},\ \bibinfo {year} {2013})\BibitemShut
  {NoStop}%
\bibitem [{Note1()}]{Note1}%
  \BibitemOpen
  \bibinfo {note} {The distance $\protect \mathcal {D}_{KS} = \protect \qopname
  \relax m{max}| F(r_k | \protect \mathcal {S} = S_i) - F(r_k | \cev {r}_{\ell
  })|$, where $F(r_k | \protect \mathcal {S} = S_i)$ and $F(r_k | \cev
  {r}_{\ell })$ are cumulative distributions of $P(r_k | \protect \mathcal {S}
  = S_i)$ and $P(r_k | \cev {r}_{\ell })$ respectively.}\BibitemShut {Stop}%
\bibitem [{\citenamefont {Miller}(1956)}]{Miller1956}%
  \BibitemOpen
  \bibfield  {author} {\bibinfo {author} {\bibfnamefont {L.~H.}\ \bibnamefont
  {Miller}},\ }\href {\doibase 10.1080/01621459.1956.10501314} {\bibfield
  {journal} {\bibinfo  {journal} {J. Am. Stat. Assoc.}\ }\textbf {\bibinfo
  {volume} {51}},\ \bibinfo {pages} {111} (\bibinfo {year} {1956})}\BibitemShut
  {NoStop}%
\bibitem [{\citenamefont {Marton}\ and\ \citenamefont
  {Shields}(1994)}]{MartonAoP}%
  \BibitemOpen
  \bibfield  {author} {\bibinfo {author} {\bibfnamefont {K.}~\bibnamefont
  {Marton}}\ and\ \bibinfo {author} {\bibfnamefont {P.~C.}\ \bibnamefont
  {Shields}},\ }\href {https://www.jstor.org/stable/2244900} {\bibfield
  {journal} {\bibinfo  {journal} {Ann. Probab.}\ }\textbf {\bibinfo {volume}
  {23}} (\bibinfo {year} {1994})}\BibitemShut {NoStop}%
\bibitem [{\citenamefont {Cover}\ and\ \citenamefont
  {Thomas}(1991)}]{CoverThomas}%
  \BibitemOpen
  \bibfield  {author} {\bibinfo {author} {\bibfnamefont {T.~M.}\ \bibnamefont
  {Cover}}\ and\ \bibinfo {author} {\bibfnamefont {J.~A.}\ \bibnamefont
  {Thomas}},\ }\href@noop {} {\emph {\bibinfo {title} {Elements of Information
  Theory}}}\ (\bibinfo  {publisher} {Wiley, New York},\ \bibinfo {year}
  {1991})\BibitemShut {NoStop}%
\bibitem [{\citenamefont {Yang}\ \emph {et~al.}(2019)\citenamefont {Yang},
  \citenamefont {Binder}, \citenamefont {Gu},\ and\ \citenamefont
  {Elliott}}]{Yang2019}%
  \BibitemOpen
  \bibfield  {author} {\bibinfo {author} {\bibfnamefont {C.}~\bibnamefont
  {Yang}}, \bibinfo {author} {\bibfnamefont {F.~C.}\ \bibnamefont {Binder}},
  \bibinfo {author} {\bibfnamefont {M.}~\bibnamefont {Gu}}, \ and\ \bibinfo
  {author} {\bibfnamefont {T.~J.}\ \bibnamefont {Elliott}},\ }\href
  {https://arxiv.org/abs/1909.08366} {\bibfield  {journal} {\bibinfo  {journal}
  {arXiv:1909.08366}\ } (\bibinfo {year} {2019})}\BibitemShut {NoStop}%
\bibitem [{\citenamefont {{Rached}}\ \emph {et~al.}(2004)\citenamefont
  {{Rached}}, \citenamefont {{Alajaji}},\ and\ \citenamefont
  {{Campbell}}}]{Rached2004}%
  \BibitemOpen
  \bibfield  {author} {\bibinfo {author} {\bibfnamefont {Z.}~\bibnamefont
  {{Rached}}}, \bibinfo {author} {\bibfnamefont {F.}~\bibnamefont {{Alajaji}}},
  \ and\ \bibinfo {author} {\bibfnamefont {L.~L.}\ \bibnamefont {{Campbell}}},\
  }\href {\doibase 10.1109/TIT.2004.826687} {\bibfield  {journal} {\bibinfo
  {journal} {IEEE Trans. Inf. Theory}\ }\textbf {\bibinfo {volume} {50}},\
  \bibinfo {pages} {917} (\bibinfo {year} {2004})}\BibitemShut {NoStop}%
\bibitem [{\citenamefont {Amari}(2016)}]{amari2016book}%
  \BibitemOpen
  \bibfield  {author} {\bibinfo {author} {\bibfnamefont {S.}~\bibnamefont
  {Amari}},\ }\href@noop {} {\emph {\bibinfo {title} {Information geometry and
  its applications}}},\ Vol.\ \bibinfo {volume} {194}\ (\bibinfo  {publisher}
  {Springer},\ \bibinfo {year} {2016})\BibitemShut {NoStop}%
\bibitem [{\citenamefont {Amari}\ \emph {et~al.}(2018)\citenamefont {Amari},
  \citenamefont {Tsuchiya},\ and\ \citenamefont {Oizumi}}]{amari2018}%
  \BibitemOpen
  \bibfield  {author} {\bibinfo {author} {\bibfnamefont {S.}~\bibnamefont
  {Amari}}, \bibinfo {author} {\bibfnamefont {N.}~\bibnamefont {Tsuchiya}}, \
  and\ \bibinfo {author} {\bibfnamefont {M.}~\bibnamefont {Oizumi}},\ }in\
  \href {\doibase 10.1007/978-3-319-97798-0_1} {\emph {\bibinfo {booktitle}
  {Information Geometry and Its Applications}}},\ \bibinfo {editor} {edited by\
  \bibinfo {editor} {\bibfnamefont {N.}~\bibnamefont {Ay}}, \bibinfo {editor}
  {\bibfnamefont {P.}~\bibnamefont {Gibilisco}}, \ and\ \bibinfo {editor}
  {\bibfnamefont {F.}~\bibnamefont {Mat{\'u}{\v{s}}}}}\ (\bibinfo  {publisher}
  {Springer International Publishing},\ \bibinfo {address} {Cham},\ \bibinfo
  {year} {2018})\ pp.\ \bibinfo {pages} {3--17}\BibitemShut {NoStop}%
\bibitem [{\citenamefont {Paulk}\ \emph {et~al.}(2013)\citenamefont {Paulk},
  \citenamefont {Zhou}, \citenamefont {Stratton}, \citenamefont {Liu},\ and\
  \citenamefont {van Swinderen}}]{Paulk2013b}%
  \BibitemOpen
  \bibfield  {author} {\bibinfo {author} {\bibfnamefont {A.~C.}\ \bibnamefont
  {Paulk}}, \bibinfo {author} {\bibfnamefont {Y.}~\bibnamefont {Zhou}},
  \bibinfo {author} {\bibfnamefont {P.}~\bibnamefont {Stratton}}, \bibinfo
  {author} {\bibfnamefont {L.}~\bibnamefont {Liu}}, \ and\ \bibinfo {author}
  {\bibfnamefont {B.}~\bibnamefont {van Swinderen}},\ }\href {\doibase
  10.1152/jn.00414.2013} {\bibfield  {journal} {\bibinfo  {journal} {J.
  Neurophysiol.}\ }\textbf {\bibinfo {volume} {110}},\ \bibinfo {pages} {1703}
  (\bibinfo {year} {2013})}\BibitemShut {NoStop}%
\bibitem [{Note2()}]{Note2}%
  \BibitemOpen
  \bibinfo {note} {$L(N) \sim 14$ only serves as a lower bound on $\lambda $,
  past which CSSR is guaranteed to return incorrect causal states for the
  neural data. In practice, this may occur at even lower memory lengths than
  this limit. We observed this effect marked by an exponential increase in the
  number of inferred causal states for $\lambda > 11$, and thus excluded these
  memory lengths from the study.}\BibitemShut {Stop}%
\bibitem [{\citenamefont {Harrison}\ \emph {et~al.}(2018)\citenamefont
  {Harrison}, \citenamefont {Donaldson}, \citenamefont {Correa-Cano},
  \citenamefont {Evans}, \citenamefont {Fisher}, \citenamefont {Goodwin},
  \citenamefont {Robinson}, \citenamefont {Hodgson},\ and\ \citenamefont
  {Inger}}]{Harrison2018lme}%
  \BibitemOpen
  \bibfield  {author} {\bibinfo {author} {\bibfnamefont {X.~A.}\ \bibnamefont
  {Harrison}}, \bibinfo {author} {\bibfnamefont {L.}~\bibnamefont {Donaldson}},
  \bibinfo {author} {\bibfnamefont {M.~E.}\ \bibnamefont {Correa-Cano}},
  \bibinfo {author} {\bibfnamefont {J.}~\bibnamefont {Evans}}, \bibinfo
  {author} {\bibfnamefont {D.~N.}\ \bibnamefont {Fisher}}, \bibinfo {author}
  {\bibfnamefont {C.~E.}\ \bibnamefont {Goodwin}}, \bibinfo {author}
  {\bibfnamefont {B.~S.}\ \bibnamefont {Robinson}}, \bibinfo {author}
  {\bibfnamefont {D.~J.}\ \bibnamefont {Hodgson}}, \ and\ \bibinfo {author}
  {\bibfnamefont {R.}~\bibnamefont {Inger}},\ }\href {\doibase
  10.7717/peerj.4794} {\bibfield  {journal} {\bibinfo  {journal} {PeerJ}\
  }\textbf {\bibinfo {volume} {6}},\ \bibinfo {pages} {e4794} (\bibinfo {year}
  {2018})}\BibitemShut {NoStop}%
\bibitem [{\citenamefont {Bates}\ \emph {et~al.}(2015)\citenamefont {Bates},
  \citenamefont {M\"achler}, \citenamefont {Bolker},\ and\ \citenamefont
  {Walker}}]{Bates2015}%
  \BibitemOpen
  \bibfield  {author} {\bibinfo {author} {\bibfnamefont {D.}~\bibnamefont
  {Bates}}, \bibinfo {author} {\bibfnamefont {M.}~\bibnamefont {M\"achler}},
  \bibinfo {author} {\bibfnamefont {B.}~\bibnamefont {Bolker}}, \ and\ \bibinfo
  {author} {\bibfnamefont {S.}~\bibnamefont {Walker}},\ }\href {\doibase
  10.18637/jss.v067.i01} {\bibfield  {journal} {\bibinfo  {journal} {J. Stat.
  Softw.}\ }\textbf {\bibinfo {volume} {67}},\ \bibinfo {pages} {1} (\bibinfo
  {year} {2015})}\BibitemShut {NoStop}%
\bibitem [{\citenamefont {{Johnson}}\ \emph {et~al.}(2010)\citenamefont
  {{Johnson}}, \citenamefont {{Crutchfield}}, \citenamefont {{Ellison}},\ and\
  \citenamefont {{McTague}}}]{Johnson2010}%
  \BibitemOpen
  \bibfield  {author} {\bibinfo {author} {\bibfnamefont {B.~D.}\ \bibnamefont
  {{Johnson}}}, \bibinfo {author} {\bibfnamefont {J.~P.}\ \bibnamefont
  {{Crutchfield}}}, \bibinfo {author} {\bibfnamefont {C.~J.}\ \bibnamefont
  {{Ellison}}}, \ and\ \bibinfo {author} {\bibfnamefont {C.~S.}\ \bibnamefont
  {{McTague}}},\ }\href {https://ui.adsabs.harvard.edu/abs/2010arXiv1011.0036J}
  {\bibfield  {journal} {\bibinfo  {journal} {arXiv:1011.0036}\ } (\bibinfo
  {year} {2010})}\BibitemShut {NoStop}%
\bibitem [{\citenamefont {Sugihara}\ \emph {et~al.}(2012)\citenamefont
  {Sugihara}, \citenamefont {May}, \citenamefont {Ye}, \citenamefont {Hsieh},
  \citenamefont {Deyle}, \citenamefont {Fogarty},\ and\ \citenamefont
  {Munch}}]{Sugihara2012}%
  \BibitemOpen
  \bibfield  {author} {\bibinfo {author} {\bibfnamefont {G.}~\bibnamefont
  {Sugihara}}, \bibinfo {author} {\bibfnamefont {R.}~\bibnamefont {May}},
  \bibinfo {author} {\bibfnamefont {H.}~\bibnamefont {Ye}}, \bibinfo {author}
  {\bibfnamefont {C.-h.}\ \bibnamefont {Hsieh}}, \bibinfo {author}
  {\bibfnamefont {E.}~\bibnamefont {Deyle}}, \bibinfo {author} {\bibfnamefont
  {M.}~\bibnamefont {Fogarty}}, \ and\ \bibinfo {author} {\bibfnamefont
  {S.}~\bibnamefont {Munch}},\ }\href {\doibase 10.1126/science.1227079}
  {\bibfield  {journal} {\bibinfo  {journal} {Science}\ }\textbf {\bibinfo
  {volume} {338}},\ \bibinfo {pages} {496} (\bibinfo {year}
  {2012})}\BibitemShut {NoStop}%
\bibitem [{\citenamefont {Tajima}\ \emph {et~al.}(2015)\citenamefont {Tajima},
  \citenamefont {Yanagawa}, \citenamefont {Fujii},\ and\ \citenamefont
  {Toyoizumi}}]{Tajima2015}%
  \BibitemOpen
  \bibfield  {author} {\bibinfo {author} {\bibfnamefont {S.}~\bibnamefont
  {Tajima}}, \bibinfo {author} {\bibfnamefont {T.}~\bibnamefont {Yanagawa}},
  \bibinfo {author} {\bibfnamefont {N.}~\bibnamefont {Fujii}}, \ and\ \bibinfo
  {author} {\bibfnamefont {T.}~\bibnamefont {Toyoizumi}},\ }\href {\doibase
  10.1371/journal.pcbi.1004537} {\bibfield  {journal} {\bibinfo  {journal}
  {PLOS Comput. Biol.}\ }\textbf {\bibinfo {volume} {11}},\ \bibinfo {pages}
  {e1004537} (\bibinfo {year} {2015})}\BibitemShut {NoStop}%
\bibitem [{\citenamefont {Breuer}\ \emph {et~al.}(2009)\citenamefont {Breuer},
  \citenamefont {Laine},\ and\ \citenamefont {Piilo}}]{BreuerPRL2009}%
  \BibitemOpen
  \bibfield  {author} {\bibinfo {author} {\bibfnamefont {H.-P.}\ \bibnamefont
  {Breuer}}, \bibinfo {author} {\bibfnamefont {E.-M.}\ \bibnamefont {Laine}}, \
  and\ \bibinfo {author} {\bibfnamefont {J.}~\bibnamefont {Piilo}},\ }\href
  {\doibase 10.1103/PhysRevLett.103.210401} {\bibfield  {journal} {\bibinfo
  {journal} {Phys. Rev. Lett.}\ }\textbf {\bibinfo {volume} {103}},\ \bibinfo
  {pages} {210401} (\bibinfo {year} {2009})}\BibitemShut {NoStop}%
\bibitem [{\citenamefont {Sarasso}\ \emph {et~al.}(2015)\citenamefont
  {Sarasso}, \citenamefont {Boly}, \citenamefont {Napolitani}, \citenamefont
  {Gosseries}, \citenamefont {Charland-Verville}, \citenamefont {Casarotto},
  \citenamefont {Rosanova}, \citenamefont {Casali}, \citenamefont {Brichant},
  \citenamefont {Boveroux}, \citenamefont {Rex}, \citenamefont {Tononi},
  \citenamefont {Laureys},\ and\ \citenamefont {Massimini}}]{Sarasso2015}%
  \BibitemOpen
  \bibfield  {author} {\bibinfo {author} {\bibfnamefont {S.}~\bibnamefont
  {Sarasso}}, \bibinfo {author} {\bibfnamefont {M.}~\bibnamefont {Boly}},
  \bibinfo {author} {\bibfnamefont {M.}~\bibnamefont {Napolitani}}, \bibinfo
  {author} {\bibfnamefont {O.}~\bibnamefont {Gosseries}}, \bibinfo {author}
  {\bibfnamefont {V.}~\bibnamefont {Charland-Verville}}, \bibinfo {author}
  {\bibfnamefont {S.}~\bibnamefont {Casarotto}}, \bibinfo {author}
  {\bibfnamefont {M.}~\bibnamefont {Rosanova}}, \bibinfo {author}
  {\bibfnamefont {A.~G.}\ \bibnamefont {Casali}}, \bibinfo {author}
  {\bibfnamefont {J.-F.}\ \bibnamefont {Brichant}}, \bibinfo {author}
  {\bibfnamefont {P.}~\bibnamefont {Boveroux}}, \bibinfo {author}
  {\bibfnamefont {S.}~\bibnamefont {Rex}}, \bibinfo {author} {\bibfnamefont
  {G.}~\bibnamefont {Tononi}}, \bibinfo {author} {\bibfnamefont
  {S.}~\bibnamefont {Laureys}}, \ and\ \bibinfo {author} {\bibfnamefont
  {M.}~\bibnamefont {Massimini}},\ }\href {\doibase
  https://doi.org/10.1016/j.cub.2015.10.014} {\bibfield  {journal} {\bibinfo
  {journal} {Curr. Biol.}\ }\textbf {\bibinfo {volume} {25}},\ \bibinfo {pages}
  {3099 } (\bibinfo {year} {2015})}\BibitemShut {NoStop}%
\bibitem [{\citenamefont {Lewis}\ \emph {et~al.}(2012)\citenamefont {Lewis},
  \citenamefont {Weiner}, \citenamefont {Mukamel}, \citenamefont {Donoghue},
  \citenamefont {Eskandar}, \citenamefont {Madsen}, \citenamefont {Anderson},
  \citenamefont {Hochberg}, \citenamefont {Cash}, \citenamefont {Brown},\ and\
  \citenamefont {Purdon}}]{Lewis2012}%
  \BibitemOpen
  \bibfield  {author} {\bibinfo {author} {\bibfnamefont {L.~D.}\ \bibnamefont
  {Lewis}}, \bibinfo {author} {\bibfnamefont {V.~S.}\ \bibnamefont {Weiner}},
  \bibinfo {author} {\bibfnamefont {E.~A.}\ \bibnamefont {Mukamel}}, \bibinfo
  {author} {\bibfnamefont {J.~A.}\ \bibnamefont {Donoghue}}, \bibinfo {author}
  {\bibfnamefont {E.~N.}\ \bibnamefont {Eskandar}}, \bibinfo {author}
  {\bibfnamefont {J.~R.}\ \bibnamefont {Madsen}}, \bibinfo {author}
  {\bibfnamefont {W.~S.}\ \bibnamefont {Anderson}}, \bibinfo {author}
  {\bibfnamefont {L.~R.}\ \bibnamefont {Hochberg}}, \bibinfo {author}
  {\bibfnamefont {S.~S.}\ \bibnamefont {Cash}}, \bibinfo {author}
  {\bibfnamefont {E.~N.}\ \bibnamefont {Brown}}, \ and\ \bibinfo {author}
  {\bibfnamefont {P.~L.}\ \bibnamefont {Purdon}},\ }\href {\doibase
  10.1073/pnas.1210907109} {\bibfield  {journal} {\bibinfo  {journal} {Proc.
  Natl. Acad. Sci.}\ }\textbf {\bibinfo {volume} {109}},\ \bibinfo {pages}
  {E3377} (\bibinfo {year} {2012})}\BibitemShut {NoStop}%
\bibitem [{\citenamefont {Raccuglia}\ \emph {et~al.}(2019)\citenamefont
  {Raccuglia}, \citenamefont {Huang}, \citenamefont {Ender}, \citenamefont
  {Heim}, \citenamefont {Laber}, \citenamefont {Su{\'a}rez-Grimalt},
  \citenamefont {Liotta}, \citenamefont {Sigrist}, \citenamefont {Geiger},\
  and\ \citenamefont {Owald}}]{raccuglia2019}%
  \BibitemOpen
  \bibfield  {author} {\bibinfo {author} {\bibfnamefont {D.}~\bibnamefont
  {Raccuglia}}, \bibinfo {author} {\bibfnamefont {S.}~\bibnamefont {Huang}},
  \bibinfo {author} {\bibfnamefont {A.}~\bibnamefont {Ender}}, \bibinfo
  {author} {\bibfnamefont {M.-M.}\ \bibnamefont {Heim}}, \bibinfo {author}
  {\bibfnamefont {D.}~\bibnamefont {Laber}}, \bibinfo {author} {\bibfnamefont
  {R.}~\bibnamefont {Su{\'a}rez-Grimalt}}, \bibinfo {author} {\bibfnamefont
  {A.}~\bibnamefont {Liotta}}, \bibinfo {author} {\bibfnamefont {S.~J.}\
  \bibnamefont {Sigrist}}, \bibinfo {author} {\bibfnamefont {J.~R.}\
  \bibnamefont {Geiger}}, \ and\ \bibinfo {author} {\bibfnamefont
  {D.}~\bibnamefont {Owald}},\ }\href {\doibase 10.1016/j.cub.2019.08.070}
  {\bibfield  {journal} {\bibinfo  {journal} {Curr. Biol.}\ }\textbf {\bibinfo
  {volume} {29}},\ \bibinfo {pages} {3611} (\bibinfo {year}
  {2019})}\BibitemShut {NoStop}%
\bibitem [{\citenamefont {Friston}(2010)}]{Friston2010}%
  \BibitemOpen
  \bibfield  {author} {\bibinfo {author} {\bibfnamefont {K.}~\bibnamefont
  {Friston}},\ }\href {\doibase 10.1038/nrn2787} {\bibfield  {journal}
  {\bibinfo  {journal} {Nat. Rev. Neurosci.}\ }\textbf {\bibinfo {volume}
  {11}},\ \bibinfo {pages} {127} (\bibinfo {year} {2010})}\BibitemShut
  {NoStop}%
\bibitem [{\citenamefont {Tani}\ and\ \citenamefont {Nolfi}(1999)}]{Tani1999}%
  \BibitemOpen
  \bibfield  {author} {\bibinfo {author} {\bibfnamefont {J.}~\bibnamefont
  {Tani}}\ and\ \bibinfo {author} {\bibfnamefont {S.}~\bibnamefont {Nolfi}},\
  }\href {\doibase https://doi.org/10.1016/S0893-6080(99)00060-X} {\bibfield
  {journal} {\bibinfo  {journal} {Neural Netw.}\ }\textbf {\bibinfo {volume}
  {12}},\ \bibinfo {pages} {1131 } (\bibinfo {year} {1999})}\BibitemShut
  {NoStop}%
\bibitem [{\citenamefont {Bastos}\ \emph {et~al.}(2012)\citenamefont {Bastos},
  \citenamefont {Usrey}, \citenamefont {Adams}, \citenamefont {Mangun},
  \citenamefont {Fries},\ and\ \citenamefont {Friston}}]{Bastos2012}%
  \BibitemOpen
  \bibfield  {author} {\bibinfo {author} {\bibfnamefont {A.~M.}\ \bibnamefont
  {Bastos}}, \bibinfo {author} {\bibfnamefont {W.~M.}\ \bibnamefont {Usrey}},
  \bibinfo {author} {\bibfnamefont {R.~A.}\ \bibnamefont {Adams}}, \bibinfo
  {author} {\bibfnamefont {G.~R.}\ \bibnamefont {Mangun}}, \bibinfo {author}
  {\bibfnamefont {P.}~\bibnamefont {Fries}}, \ and\ \bibinfo {author}
  {\bibfnamefont {K.~J.}\ \bibnamefont {Friston}},\ }\bibfield  {booktitle}
  {\emph {\bibinfo {booktitle} {Neuron}},\ }\href {\doibase
  10.1016/j.neuron.2012.10.038} {\bibfield  {journal} {\bibinfo  {journal}
  {Neuron}\ }\textbf {\bibinfo {volume} {76}},\ \bibinfo {pages} {695}
  (\bibinfo {year} {2012})}\BibitemShut {NoStop}%
\bibitem [{\citenamefont {Cofr\'{e}}\ \emph {et~al.}(2019)\citenamefont
  {Cofr\'{e}}, \citenamefont {Videla},\ and\ \citenamefont
  {Rosas}}]{cofre2019}%
  \BibitemOpen
  \bibfield  {author} {\bibinfo {author} {\bibfnamefont {R.}~\bibnamefont
  {Cofr\'{e}}}, \bibinfo {author} {\bibfnamefont {L.}~\bibnamefont {Videla}}, \
  and\ \bibinfo {author} {\bibfnamefont {F.}~\bibnamefont {Rosas}},\ }\href
  {\doibase 10.3390/e21090884} {\bibfield  {journal} {\bibinfo  {journal}
  {Entropy}\ }\textbf {\bibinfo {volume} {21}},\ \bibinfo {pages} {e21090884}
  (\bibinfo {year} {2019})}\BibitemShut {NoStop}%
\bibitem [{\citenamefont {Cofr\'{e}}\ and\ \citenamefont
  {Maldonado}(2018)}]{cofre2018}%
  \BibitemOpen
  \bibfield  {author} {\bibinfo {author} {\bibfnamefont {R.}~\bibnamefont
  {Cofr\'{e}}}\ and\ \bibinfo {author} {\bibfnamefont {C.}~\bibnamefont
  {Maldonado}},\ }\href {\doibase 10.3390/e20010034} {\bibfield  {journal}
  {\bibinfo  {journal} {Entropy}\ }\textbf {\bibinfo {volume} {20}},\ \bibinfo
  {pages} {e20010034} (\bibinfo {year} {2018})}\BibitemShut {NoStop}%
\bibitem [{\citenamefont {Hasson}\ \emph {et~al.}(2008)\citenamefont {Hasson},
  \citenamefont {Yang}, \citenamefont {Vallines}, \citenamefont {Heeger},\ and\
  \citenamefont {Rubin}}]{hasson2008}%
  \BibitemOpen
  \bibfield  {author} {\bibinfo {author} {\bibfnamefont {U.}~\bibnamefont
  {Hasson}}, \bibinfo {author} {\bibfnamefont {E.}~\bibnamefont {Yang}},
  \bibinfo {author} {\bibfnamefont {I.}~\bibnamefont {Vallines}}, \bibinfo
  {author} {\bibfnamefont {D.~J.}\ \bibnamefont {Heeger}}, \ and\ \bibinfo
  {author} {\bibfnamefont {N.}~\bibnamefont {Rubin}},\ }\href {\doibase
  10.1523/JNEUROSCI.5487-07.2008} {\bibfield  {journal} {\bibinfo  {journal}
  {J. Neurosci.}\ }\textbf {\bibinfo {volume} {28}},\ \bibinfo {pages} {2539}
  (\bibinfo {year} {2008})}\BibitemShut {NoStop}%
\bibitem [{\citenamefont {Diba}\ and\ \citenamefont
  {Buzs\'{a}ki}(2007)}]{Diba2007}%
  \BibitemOpen
  \bibfield  {author} {\bibinfo {author} {\bibfnamefont {K.}~\bibnamefont
  {Diba}}\ and\ \bibinfo {author} {\bibfnamefont {G.}~\bibnamefont
  {Buzs\'{a}ki}},\ }\href {\doibase 10.1038/nn1961} {\bibfield  {journal}
  {\bibinfo  {journal} {Nat. Neurosci.}\ }\textbf {\bibinfo {volume} {10}},\
  \bibinfo {pages} {1241} (\bibinfo {year} {2007})}\BibitemShut {NoStop}%
\bibitem [{\citenamefont {Edmonds}(1997)}]{Edmonds1997}%
  \BibitemOpen
  \bibfield  {author} {\bibinfo {author} {\bibfnamefont {B.~H.}\ \bibnamefont
  {Edmonds}},\ }\href {http://bruce.edmonds.name/combib/} {\enquote {\bibinfo
  {title} {Hypertext bibliography of measures of complexity},}\ } (\bibinfo
  {year} {1997})\BibitemShut {NoStop}%
\bibitem [{\citenamefont {Valdez}\ \emph {et~al.}(2017)\citenamefont {Valdez},
  \citenamefont {Jaschke}, \citenamefont {Vargas},\ and\ \citenamefont
  {Carr}}]{PhysRevLett.119.225301}%
  \BibitemOpen
  \bibfield  {author} {\bibinfo {author} {\bibfnamefont {M.~A.}\ \bibnamefont
  {Valdez}}, \bibinfo {author} {\bibfnamefont {D.}~\bibnamefont {Jaschke}},
  \bibinfo {author} {\bibfnamefont {D.~L.}\ \bibnamefont {Vargas}}, \ and\
  \bibinfo {author} {\bibfnamefont {L.~D.}\ \bibnamefont {Carr}},\ }\href
  {\doibase 10.1103/PhysRevLett.119.225301} {\bibfield  {journal} {\bibinfo
  {journal} {Phys. Rev. Lett.}\ }\textbf {\bibinfo {volume} {119}},\ \bibinfo
  {pages} {225301} (\bibinfo {year} {2017})}\BibitemShut {NoStop}%
\bibitem [{\citenamefont {Sundar}\ \emph {et~al.}(2018)\citenamefont {Sundar},
  \citenamefont {Valdez}, \citenamefont {Carr},\ and\ \citenamefont
  {Hazzard}}]{PhysRevA.97.052320}%
  \BibitemOpen
  \bibfield  {author} {\bibinfo {author} {\bibfnamefont {B.}~\bibnamefont
  {Sundar}}, \bibinfo {author} {\bibfnamefont {M.~A.}\ \bibnamefont {Valdez}},
  \bibinfo {author} {\bibfnamefont {L.~D.}\ \bibnamefont {Carr}}, \ and\
  \bibinfo {author} {\bibfnamefont {K.~R.~A.}\ \bibnamefont {Hazzard}},\ }\href
  {\doibase 10.1103/PhysRevA.97.052320} {\bibfield  {journal} {\bibinfo
  {journal} {Phys. Rev. A}\ }\textbf {\bibinfo {volume} {97}},\ \bibinfo
  {pages} {052320} (\bibinfo {year} {2018})}\BibitemShut {NoStop}%
\bibitem [{\citenamefont {Zanardi}\ \emph {et~al.}(2018)\citenamefont
  {Zanardi}, \citenamefont {Tomka},\ and\ \citenamefont {Venuti}}]{QIIT}%
  \BibitemOpen
  \bibfield  {author} {\bibinfo {author} {\bibfnamefont {P.}~\bibnamefont
  {Zanardi}}, \bibinfo {author} {\bibfnamefont {M.}~\bibnamefont {Tomka}}, \
  and\ \bibinfo {author} {\bibfnamefont {L.~C.}\ \bibnamefont {Venuti}},\
  }\href {https://www.arxiv.org/abs1806.01421} {\bibfield  {journal} {\bibinfo
  {journal} {arXiv:1806.01421}\ } (\bibinfo {year} {2018})}\BibitemShut
  {NoStop}%
\bibitem [{\citenamefont {Feldman}\ and\ \citenamefont
  {Crutchfield}(1998)}]{whyCmu}%
  \BibitemOpen
  \bibfield  {author} {\bibinfo {author} {\bibfnamefont {D.~P.}\ \bibnamefont
  {Feldman}}\ and\ \bibinfo {author} {\bibfnamefont {J.~P.}\ \bibnamefont
  {Crutchfield}},\ }\href {\doibase 10.1016/S0375-9601(97)00855-4} {\bibfield
  {journal} {\bibinfo  {journal} {Phys. Lett. A}\ }\textbf {\bibinfo {volume}
  {238}},\ \bibinfo {pages} {244} (\bibinfo {year} {1998})}\BibitemShut
  {NoStop}%
\bibitem [{\citenamefont {Schartner}\ \emph {et~al.}(2015)\citenamefont
  {Schartner}, \citenamefont {Seth}, \citenamefont {Noirhomme}, \citenamefont
  {Boly}, \citenamefont {Bruno}, \citenamefont {Laureys},\ and\ \citenamefont
  {Barrett}}]{schartner2015}%
  \BibitemOpen
  \bibfield  {author} {\bibinfo {author} {\bibfnamefont {M.}~\bibnamefont
  {Schartner}}, \bibinfo {author} {\bibfnamefont {A.}~\bibnamefont {Seth}},
  \bibinfo {author} {\bibfnamefont {Q.}~\bibnamefont {Noirhomme}}, \bibinfo
  {author} {\bibfnamefont {M.}~\bibnamefont {Boly}}, \bibinfo {author}
  {\bibfnamefont {M.-A.}\ \bibnamefont {Bruno}}, \bibinfo {author}
  {\bibfnamefont {S.}~\bibnamefont {Laureys}}, \ and\ \bibinfo {author}
  {\bibfnamefont {A.}~\bibnamefont {Barrett}},\ }\href {\doibase
  10.1371/journal.pone.0133532} {\bibfield  {journal} {\bibinfo  {journal}
  {PLOS One}\ }\textbf {\bibinfo {volume} {10}},\ \bibinfo {pages} {1}
  (\bibinfo {year} {2015})}\BibitemShut {NoStop}%
\bibitem [{\citenamefont {Ellison}\ \emph {et~al.}(2011)\citenamefont
  {Ellison}, \citenamefont {Mahoney}, \citenamefont {James}, \citenamefont
  {Crutchfield},\ and\ \citenamefont {Reichardt}}]{Ellison2011}%
  \BibitemOpen
  \bibfield  {author} {\bibinfo {author} {\bibfnamefont {C.~J.}\ \bibnamefont
  {Ellison}}, \bibinfo {author} {\bibfnamefont {J.~R.}\ \bibnamefont
  {Mahoney}}, \bibinfo {author} {\bibfnamefont {R.~G.}\ \bibnamefont {James}},
  \bibinfo {author} {\bibfnamefont {J.~P.}\ \bibnamefont {Crutchfield}}, \ and\
  \bibinfo {author} {\bibfnamefont {J.}~\bibnamefont {Reichardt}},\ }\href
  {\doibase 10.1063/1.3637490} {\bibfield  {journal} {\bibinfo  {journal}
  {Chaos}\ }\textbf {\bibinfo {volume} {21}},\ \bibinfo {pages} {037107}
  (\bibinfo {year} {2011})}\BibitemShut {NoStop}%
\end{thebibliography}%
\end{document}